\documentclass[11pt,a4paper]{article}
 
\usepackage{graphicx}
\usepackage{epstopdf}
\usepackage{jcappub}

\usepackage{amsmath,amsthm,latexsym,amssymb,amsfonts,epsfig}
\usepackage{fontenc,dsfont,layout}
\usepackage{hyperref}
\usepackage{psfrag}
\usepackage[cp1250]{inputenc} 

\usepackage[usenames,dvipsnames]{xcolor}

\numberwithin{equation}{section}

\title{Inflationary scenarios in Starobinsky model with higher order corrections}
\author[a]{Micha{\l}~Artymowski}
\author[b]{Zygmunt~Lalak}
\author[bc]{Marek~Lewicki}

\affiliation[a]{Institute of Physics, Jagiellonian University\\
{\L}ojasiewicza 11, 30-348 Krak{\'o}w, Poland}
\affiliation[b]{Institute of Theoretical Physics, Faculty of Physics, University of Warsaw\\
 ul. Pasteura 5, 02-093 Warsaw, Poland}
\affiliation[c]{Michigan Center for Theoretical Physics, University of Michigan\\
 Ann Arbor MI 48109, USA} 

\emailAdd{Michal.Artymowski@uj.edu.pl}
\emailAdd{Zygmunt.Lalak@fuw.edu.pl}
\emailAdd{Marek.Lewicki@fuw.edu.pl}

\abstract{We consider the Starobinsky inflation with a set of higher order corrections parametrised by two real coefficients $\lambda_1\, , \lambda_2$. In the Einstein frame we have found a potential with the Starobinsky plateau, steep slope and possibly with an additional minimum, local maximum or a saddle point. We have identified three types of inflationary behaviour that may be generated in this model: i) inflation on the plateau, ii) at the local maximum (topological inflation), iii) at the saddle point. We have found limits on parameters $\lambda_i$ and initial conditions at the Planck scale which enable successful inflation and disable eternal inflation at the plateau. We have checked that the local minimum away from the GR vacuum is stable and that the field cannot leave it  neither via quantum tunnelling nor via thermal corrections.}


\begin{document}
\maketitle

\section{Introduction}

The Starobinsky inflation \cite{Starobinsky:1980te} is an $f(R)=R+R^2/6M^2$ theory and  together with the original Guth model is the first model of cosmic inflation. It can be expressed as a Brans-Dicke theory with the Jordan frame potential proportional to $(\varphi-1)^2$. After 35 years from its creation it is consistent with the recent CMB data \cite{Ade:2013uln} and it has developed many generalisations   \cite{DeFelice:2010aj,Codello:2014sua,Ben-Dayan:2014isa,Artymowski:2014gea,Artymowski:2014nva,Sebastiani:2013eqa,Motohashi:2014tra}. Its predictions (up to the $N^{-2}$ terms, where $N$ is the number of e-folds until the end of inflation) coincide with the so-called Higgs inflation, where inflation is generated by the scalar field with $V\propto \lambda\phi^4$ and with non-minimal coupling to gravity of the form $\xi\phi^2 R$. 
\\*

Recently it has been shown \cite{Kallosh:2013tua,Kallosh:2013yoa,Kallosh:2014rga,Kallosh:2014laa,Galante:2014ifa} that the Starobinsky model is a part of an attractor of results on the $(n_s,r)$ plane of several inflationary models based on theories of modified gravity. Its potential (as well as the potential of the Higgs inflation) may be modified by the higher order corrections, which take the form of the additional higher powers of Starobinsky potential in the action \cite{Broy:2014sia,Kamada:2014gma}. The higher order terms may generate a steep slope of the Einstein frame potential  which limits the Starobinsky plateau and the amount of e-folds that can be generated during inflation. In this paper we want to investigate those generalisations, especially in the context of possible existence of local minima and maxima of the potential and their influence on the evolution of the inflation, existence of the successful inflation and late-time evolution of the Universe. 
\\*

The structure of this paper is as follows. In the Sec. \ref{sec:Infl} we analyse the evolution of inflaton and space time for the certain generalization of the Starobinsky potential. In the Sec. \ref{sec:infl2} we  investigate several types of inflation that could be generated by considered potentials. In the Sec. \ref{sec:Temp} we include non-zero temperature corrections. In the Sec. \ref{sec:Tunnel} we discuss the possibility of quantum tunnelling from meta-stable vacua of the Einstein frame potential. Finally, we conclude in the Sec. \ref{sec:concl}. 
\\*
In the following  we use the convention $8\pi G = M_{pl}^{-2} = 1$, where $M_{p}\sim 2\times 10^{18}GeV$ is the reduced Planck mass.

\section{Generalisation of the Starobinsky inflation} \label{sec:Infl}

\subsection{Jordan frame analysis}

Let us consider a Brans-Dicke theory in the flat FRW space-time with the metric tensor of the form $ds^2 = -dt^2 + a(t)^2(d\vec{x})^2$. Then the Jordan frame action is of the form
\begin{equation}
S =  \int d^4x \sqrt{|g|}\left[\varphi R - \frac{\omega}{2\varphi}(\nabla\varphi)^2 - U(\varphi)\right] + S_{\text{m}}\, , \label{eq:actionJ}
\end{equation}
where $\omega = const$ and $S_{\text{m}}$ is the action of matter fields. From now on we will assume that $\omega = 0$, which makes the model equivalent to the $f(R)$ theory. Then, for the homogeneous field $\varphi$ the field's equation of motion and the first Friedmann equation become \cite{DeFelice:2010aj}
\begin{eqnarray}
\ddot{\varphi} + 3H\dot{\varphi} + \frac{2}{3}(\varphi U_\varphi - 2U) &=& \frac{1}{3}\left(\rho_M - 3P_M\right)\ ,\label{eq:motionBD}\\
3\left(H + \frac{\dot{\varphi}}{2\varphi}\right)^2 &=& \frac{3}{4}\left(\frac{\dot{\varphi}}{\varphi}\right)^2 + \frac{U}{\varphi} + \frac{\rho_M}{\varphi}\, , \label{eq:FriedBD}\\
\dot{\rho}_M + 3H(\rho_M + P_M) &=& 0\, , \label{eq:cont}
\end{eqnarray}
where $U_\varphi:=\frac{dU}{d\varphi}$ and $\rho_M$ and $P_M$ are energy density and pressure of matter fields respectively \footnote{In this paper we refer as matter fields to all perfect fluid components of the energy-stress tensor, like dust, radiation or scalar fields.}. Let us note that for $\varphi = 1$ one recovers the general relativity (GR). Thus the $\varphi=1$ will be denoted as the GR vacuum.
\\*

The Starobinsky inflation is a theory of cosmic inflation based on the $f(R) = R + R^2/6M^2$ Lagrangian, which can be generalized to Brans-Dicke theory with general value of $\omega$. The Jordan frame potential of the Starobinsky model is following
\begin{equation}
U_{S} = \frac{3}{4}M^2\left(\varphi-1\right)^2 \, , \label{eq:StarobinskyP}
\end{equation}
where $M$ is a mass parameter, whose value comes from the normalisation of the primordial inhomogeneities. For $\omega = 0$ one finds $M\simeq 1.5\times 10^{-5}$. The Starobinsky potential is presented at the Fig. \ref{fig:V}. In this paper we consider the extension of this model motivated by the Ref. \cite{Broy:2014sia}, namely
\begin{equation}
U = U_S\left(1 + \lambda_1 \frac{U_S}{M_{p}^4} + \lambda_2 \frac{U_S^2}{M_p^8}\right) \, , \label{eq:StarobinskyGeneral}
\end{equation}
where $\lambda_1$, $\lambda_2$ are numerical coefficients. In order to avoid $U\to-\infty$ for $\varphi \to \infty$ we assume that $\lambda_2>0$. The sign of $\lambda_1$ is undetermined. The motivation for the existence of the higher order terms is as follows: we assume that the Jordan frame field is a singlet of some theory beyond the Planck scale. Integrating out the heavy degrees of freedom gives $M_p$ in the denominator of the higher order terms in the potential. In addition, we require that the GR vacuum does exist in order to restore classical gravity.

\subsection*{The model as an $f(R)$ theory}

As mentioned previously one can define the $f(R)$ function (where $R$ is the Ricci scalar) for which
\begin{equation}
\varphi = F(R) := \frac{df}{dR}\, , \qquad U(\varphi) = \frac{1}{2}(RF - f) \, ,\qquad U_\varphi = \frac{R}{2} \, .
\end{equation}
Thus, from $\varphi(R)$ one can reconstruct $f(R)$ by $f = \int \varphi(R) dR$. For $U = U_S$ one finds $f(R) = R + \frac{R^2}{6M^2}$. For the general values of $\lambda_2$ one cannot solve $U_\varphi = R/2$ to obtain $\varphi = \varphi(R) = F(R)$, because there is no general solution of the fifth order equation. In such a case we cant write an explicit form of $f(R)$. On the other hand, for $\lambda_2 = 0$ and $\lambda_1=\lambda$ one finds 
\begin{eqnarray}
\varphi(R) &=& F(R) = \frac{1}{6} \left(6-\frac{2}{M^2 A}+\frac{A}{ \lambda }\right) \, ,\\
f(R) &=& \frac{1}{48\text{  }\lambda ^3}\left(4 \lambda ^2 (1+12 R \lambda )-\left(B-6 R \lambda ^2\right) A-\left(M^2 \lambda +3R B\right) A^2\right)\, ,
\end{eqnarray}
where $A = \left(6R \lambda ^2+2 \sqrt{\lambda ^3\left(2 M^2+9 R^2 \lambda \right)}\right)^{1/3}M^{-4/3}$ and $B = \sqrt{\lambda ^3 \left(2 M^2+9 R^2 \lambda \right)}-3 R \lambda ^2$. For $\lambda\to 0$ one recovers the Starobinsky model. Considering a general value of $\omega$ shall be the next step in our future analysis.

\subsection{Einstein frame analysis} \label{sec:EinAn}

The gravitational part of the action may obtain its canonical (minimally coupled to $\varphi$) form after transformation to the Einstein frame. Let us assume that $\varphi>0$. Then for the Einstein frame metric tensor
\begin{equation}
\tilde{g}_{\mu\nu}=\varphi g_{\mu\nu}\, , \qquad d\tilde{t}=\sqrt{\varphi}dt\, ,\qquad\tilde{a} = \sqrt{\varphi}a
\end{equation}
one obtains the action of the form of 
\begin{equation}
S[\tilde{g}_{\mu\nu},\varphi] = \int d^4\tilde{x} \sqrt{-\tilde{g}}\left[ \frac{1}{2}\tilde{R} - \frac{3}{4}\left( \frac{\tilde{\nabla}\varphi}{\varphi} \right)^2 - \frac{U(\varphi)}{\varphi^2} \right] + S_m[\tilde{g}_{\mu\nu},\varphi]\, ,
\end{equation}
where $\tilde{\nabla}$ is the derivative with respect to the Einstein frame coordinates. Matter fields are now explicitly coupled to $\varphi$ due to the fact that $d^4x\sqrt{-g} = \varphi^{-2}d^4\tilde{x}\sqrt{-\tilde{g}}$. In order to obtain the canonical kinetic term for $\varphi$ let us use the Einstein frame scalar field $\phi$ 
\begin{equation}
\phi = \sqrt{\frac{3}{2}}\log\varphi \, , \qquad \varphi = \exp\left(\sqrt{\frac{2}{3}}\phi\right) \,  .
\end{equation}
The GR vacuum appears at $\phi=0$. The action in terms of $\tilde{g}_{\mu\nu}$ and $\phi$ looks as follows
\begin{equation}
S = \int d^4x \sqrt{-\tilde{g}}\left[ \frac{1}{2}\tilde{R} - \frac{1}{2}\left( \tilde{\nabla}\phi \right)^2 - V(\phi)\right] + S_m[\tilde{g}_{\mu\nu},\varphi,\ldots] \, ,
\end{equation}
where
\begin{equation}
V = \left.\frac{U(\varphi)}{\varphi^2}\right|_{\varphi=\varphi(\phi)}
\end{equation}
and $\tilde{R}$ is the Ricci scalar of $\tilde{g}_{\mu\nu}$. In this section we will assume that the space-time may be described by the flat FRW metric tensor. Let us define the Einstein frame Hubble parameter as
\begin{equation}
\mathcal{H}:=\frac{\tilde{a}'}{\tilde{a}}\, ,\qquad \text{where} \qquad \tilde{a}' := \frac{d\tilde{a}}{d\tilde{t}} \, . \label{eq:HEin}
\end{equation}
Then for $\rho_M=P_M=0$ the first Friedmann equation and the equation of motion of $\phi$ are following
\begin{eqnarray}
3\mathcal{H}^2 = \frac{1}{2}\phi'^2 + V(\phi) \, , \label{eq:FriedEin} \\
\phi'' + 3\mathcal{H}\phi' + V_\phi = 0 \, ,\label{eq:EOMEin}
\end{eqnarray}
where $V_\phi = \frac{dV}{d\phi}$.

\subsection*{Minima and maxima of the Einstein frame potential}

\begin{figure}[h]
\centering
\includegraphics[height=5.1cm]{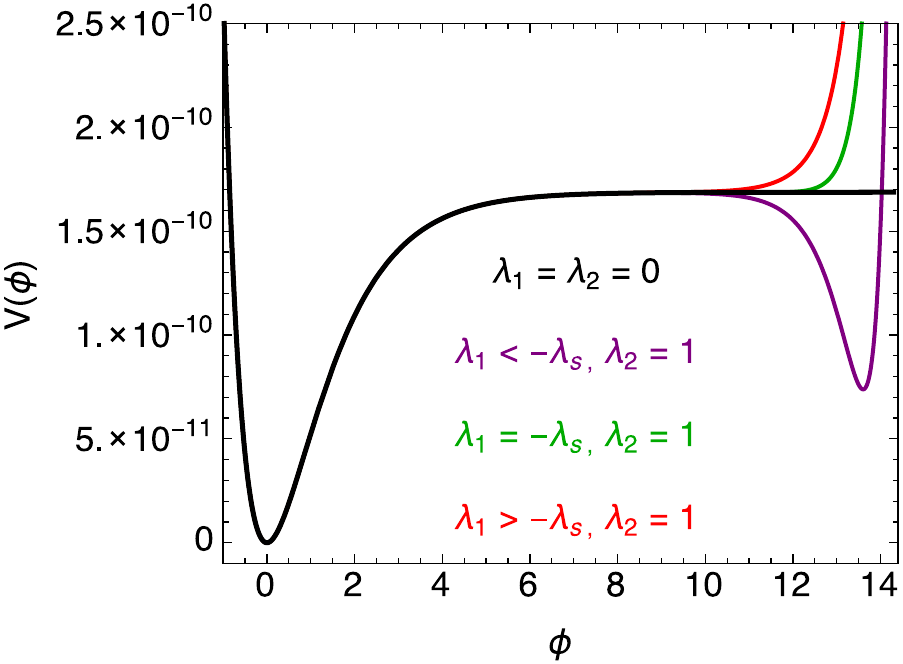}
\hspace{0.5cm}
\includegraphics[height=5.1cm]{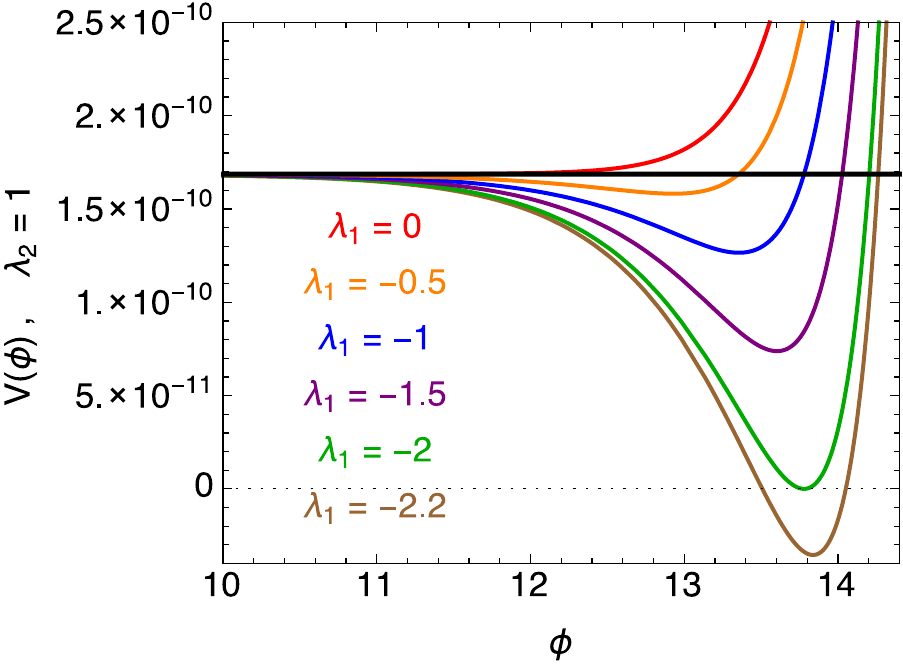}
\caption{\it The Einstein frame Starobinsky potential (thick black line) and its generalisation from the Eq. (\ref{eq:StarobinskyGeneral}) for $M=1.5\times 10^{-5}$ and for different values of $\lambda_1$ and $\lambda_2$. Left panel: The potential with minimum, saddle point and without stationary points for $\phi>0$. Right panel: The depth of the potential for different values of $\lambda_1$. Local minima with $V_{\min}>0$ may generate the de Sitter solution without  graceful exit.}
\label{fig:V}
\end{figure}

\begin{figure}[h]
\centering
\includegraphics[height=5.1cm]{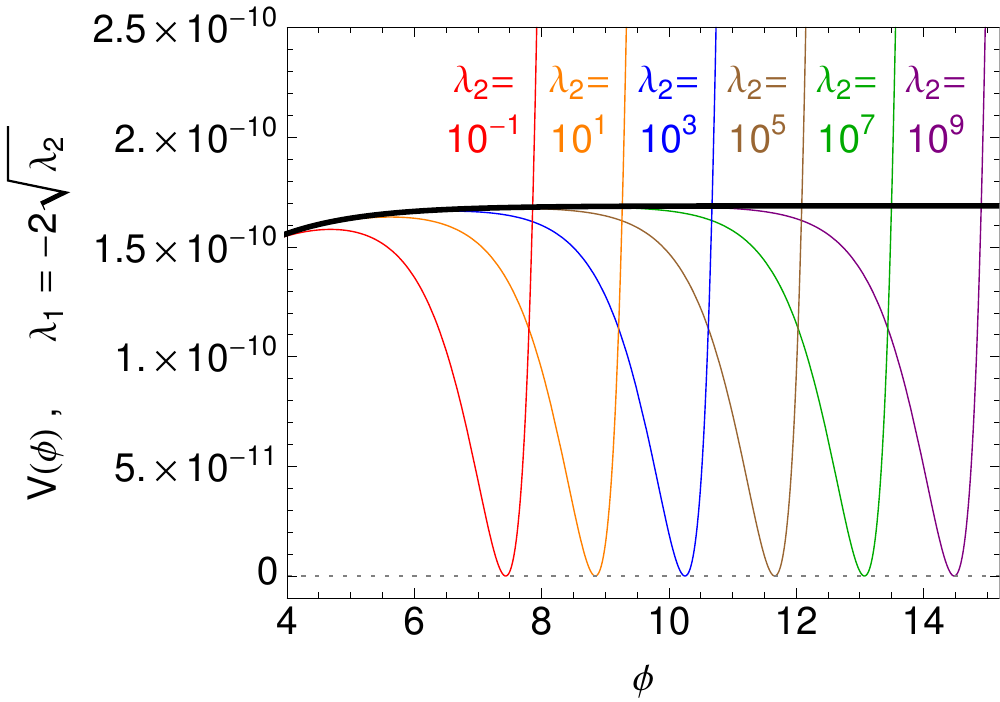}
\hspace{0.5cm}
\includegraphics[height=5.1cm]{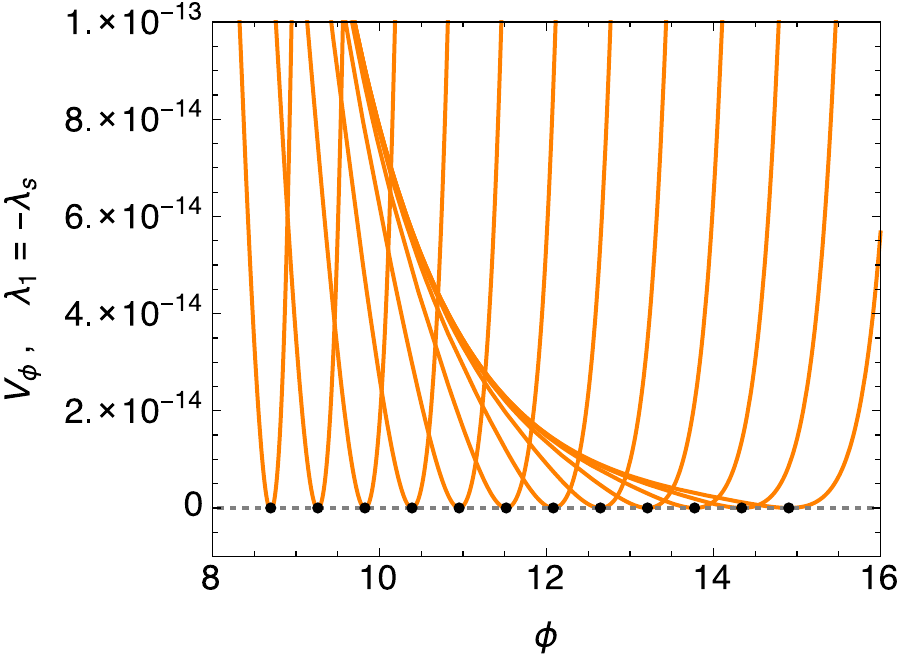}
\caption{\it Left Panel: For fixed relation between $\lambda_1$ and $\sqrt{\lambda_2}$ one finds the same value of $V_{\min}$ for all $\lambda_2 $. Right panel: $V_{\phi}$ for $\lambda_1 = -\lambda_s$. We have assumed $\lambda_2= 10^k$, where $k\in\{-7,-6,\ldots,3,4\}$ (smallest/biggest $\lambda_2$ are the right/left orange curves). Black dots represent $\phi = \phi_s$. Both $\lambda_s$ and $\phi_s$ are taken from Eq. (\ref{eq:conditionlambda}) and they appear to be very accurate. One finds $V_\phi(\phi_s) \sim \lambda_2^{2/5}M^{18/5}\lll 1$.}
\label{fig:V1}
\end{figure}

The Einstein frame potential obtained from the Eq. (\ref{eq:StarobinskyGeneral}) has following features: For $\lambda_1,\lambda_2 > 0$ one obtains the Starobinsky plateau, which ends with a steep slope when one of the higher order terms starts to dominate. In such a case $V$ has no maximum and the only minimum appears at $\phi = 0$. Thus $\phi$ ends its evolution at $\phi=0$ for any set of initial conditions. For $\lambda_1<0$ the potential has a minimum at $\phi=0$ and may have a local maximum at $\phi_{\max}>0$ and a minimum between the plateau and the slope. The last minimum will be denoted as $\phi_{\min}$ (where $\phi_{\min}>\phi_{\max}$) and its depth will depend on the relation between $\lambda_1$ and $\lambda_2$, so $\phi_{\min}$ can be a local or global minimum. Potentials with minimum at $\phi=\phi_{\min}$ are presented at Fig. \ref{fig:V} and \ref{fig:V1}
\\*

\begin{figure}[h]
\centering
\includegraphics[height=5.8cm]{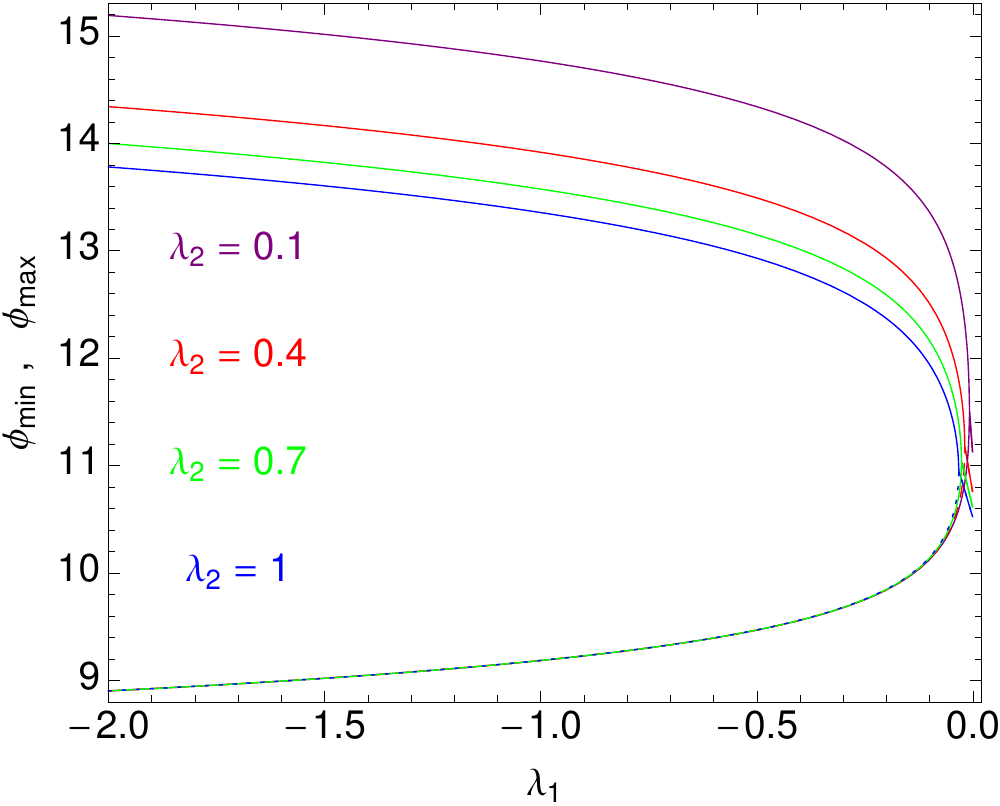}
\caption{\it The plot show numerical results for $\phi_{\min}$ and $\phi_{\max}$ (solid and dashed lines respectively) as a function of $\lambda_1$ for several different values of $\lambda_2$. Not surprisingly $\phi_{\max}$ does not depend on $\lambda_2$, which dominates the potential for bigger values of $\phi$. Both $\phi_{\min}$ and $\phi_{\max}$ meet at $\phi_s$ when the $|\lambda_1|>\lambda_s$ is violated. Those results are perfectly consistent with analytical analysis of $\phi_{\min}$ and $\phi_{\max}$ (see Eq. (\ref{eq:phimin},\ref{eq:phimax})) besides the narrow region of very small $\lambda_1$, {where} the $\lambda_2$ term shall also be taken into account in the Eq. (\ref{eq:phimax}).} 
\label{fig:phiMinMax}
\end{figure}

Let us assume that $\lambda_1<0$ and $\lambda_2>0$. In order to find an extreme value of $V$ one needs to solve the equation $V_\phi = 0$. For $\phi$ of order of a few $M_{p}$ (which is the case for $\phi_{\min}$ or $\phi_{\max}$) one finds $\exp(\sqrt{2/3}\,\phi)\gg 1$, so the condition $V_\phi = 0$ simplifies into
 \begin{equation}
8+6 e^{3\sqrt{\frac{2}{3}} \phi } M^2 \lambda _1+9 e^{5 \sqrt{\frac{2}{3}} \phi } M^4 \lambda _2= 0 \, .\label{eq:VphiZero}
\end{equation}
This equation may have solutions as long as the $\lambda_1$ can dominate other terms for some values of $\phi$. In such a case, for $\lambda_1<0$, the minimum (maximum) or the saddle point exists for $V_{\phi\phi}<0$ ($V_{\phi\phi}>0$) or $V_{\phi\phi}=0$ respectively. Let us consider the latter case in which $V_\phi$ has a saddle point at some $\phi = \phi_s$, but does not have any maxima or minima besides the one in $\phi = 0$. Then, assuming $\exp(\sqrt{2/3}\,\phi)\gg1$ and $\lambda_1 \gg M^2 \lambda_2$ one finds that the condition $V_{\phi\phi} = 0$ simplifies into
\begin{equation}
-2 + 3 \lambda _1 M^2 e^{3 \sqrt{\frac{2}{3}} \phi } + 9 \lambda _2 M^4 e^{5 \sqrt{\frac{2}{3}} \phi } =0 \label{eq:condSaddle}
\end{equation}
at $\phi = \phi_s$. Then from Eq. (\ref{eq:VphiZero},\ref{eq:condSaddle}) one finds
\begin{equation}
\phi_s \simeq \frac{1}{\sqrt{6}}\log\left(-\frac{10}{3 \lambda _1 M^2}\right) \, ,\quad \lambda_1 = -\lambda_s \simeq  - \frac{5 \lambda _2^{3/5} M^{2/5}}{2^{1/5} 3^{2/5}} \, , \quad V_s = \frac{3}{4} M^2 \left(1-\frac{15}{8}\left(24M^4\lambda _2\right){}^{1/5}\right)\, , \label{eq:conditionlambda}
\end{equation}
where $\lambda_s$ fixes the relation between $\lambda_1$ and $\lambda_2$ in order to provide the existence of a saddle point and $V_s = V(\phi_s)$. Let us note that for $\phi \simeq \phi_s$ one obtains a saddle point inflation, which in principle could significantly decrease the scale of inflation. The $v_\phi$ for $\lambda_1 = -\lambda_s$ is presented at the Fig. \ref{fig:V1}. For $|\lambda_1| < \lambda_s$ the $\lambda_1$ term is always subdominant compared to the other terms in $V(\phi)$ and it can be neglected in the analysis. For $\lambda_1>-\lambda_s$ the potential has no stationary points besides the minimum at $\phi = 0$. Lets us assume that $\lambda_1<-\lambda_s$, which means that $\phi_{\min}$ and $\phi_{\max}$ do exist. At the minimum $\exp(\sqrt{2/3}\,\phi_{\min}) \sim \mathcal{O}(M^{-1})$, so the first term in the Eq. (\ref{eq:VphiZero}) is negligible. Thus, for $|\lambda_1|,\lambda_2 < M^{-1}$ one finds
\begin{equation}
\phi _{\min } \simeq \frac{1}{2}\sqrt{\frac{3}{2}}\log \left(-\frac{2}{3} \frac{\lambda _1}{M^2 \lambda _2 }\right)\, . \label{eq:phimin}
\end{equation}
Assuming $\lambda_1,\lambda_2 < M^{-1}$ one finds the following value of $V_{\min}$
\begin{equation}
V_{\min} := V(\phi_{\min}) \simeq \frac{3 M^2 \left(4 \lambda _2-\lambda _1^2\right)}{16 \lambda _2} \, . \label{eq:Vmin}
\end{equation}
Thus $V_{\min}\geq0 \Leftrightarrow \lambda_1 \in [-2\sqrt{\lambda_2},0)$. If the field oscillates around $\phi_{\min}$ it has a mass defined by
\begin{equation}
m_{\phi}^2 := V_{\phi\phi}(\phi = \phi_{\min}) \simeq -\frac{\lambda _2^2 M^2}{\lambda _1}\, . \label{eq:Mphi}
\end{equation}
One can apply the same procedure to calculate $\phi_{\max}$. At the plateau $1\ll \exp(\sqrt{2/3}\,\phi)\ll M^{-1}$, which makes the third term of the Eq. (\ref{eq:VphiZero}) negligible. Thus $\phi_{\max}$ and $V_{\max} := V(\phi_{\max})$
\begin{equation}
\phi_{\max} \simeq \sqrt{\frac{1}{6}}\log \left( \frac{-4}{3M^2 \lambda _1}\right) \, , \qquad V_{\max} \simeq \frac{3}{4} M^2 \left(1-\frac{3}{2}\left(-6M^2 \lambda _1\right)^{1/3}\right) \, . \label{eq:phimax}
\end{equation}
Let us stress once again that $\phi_{\max}$ is only a local maximum of the potential. The values of $\phi_{\min}$ and $\phi_{\max}$ are plotted at the Fig. \ref{fig:phiMinMax}. One could argue that typical values of $\lambda_1$ and $\lambda_2$ are of order of unity, but in general they may be much bigger. The only limitation is that the steep slope shall begin for $\phi>5.5$, so the plateau is long enough to generate successful inflation. As we will show this requirement gives $|\lambda_1|\lesssim M^{-1}$ and $\lambda_2\lesssim M^{-2}$.

\subsection*{Evolution of $\phi$}
\label{sec:evolofphi}

The evolution of the field is as follows: the field starts its evolution at $\phi_0>\phi_{\min}$ and rolls down towards $\phi_{\min}$. If its kinetic energy is sufficiently big it rolls up to the plateau and generates Starobinsky-like inflation. The length of the plateau is typically around $\phi_{i}-\phi_{f}\sim 10$, where $\phi_{i}$ and $\phi_{f}$ are the beginning and the end of the plateau respectively. This is more than enough, since in Starobinsky inflation one needs $\phi_{i}\simeq 4.7$ in order to generate $60$ e-folds of inflation. Another interesting issue is that around the middle of the plateau the potential starts to decrease with $\phi$, due to the existence of local maximum at $\phi_{\max}$. As we will show this could lead to eternal inflation \cite{Guth:2007ng} at $\phi_{\max}$. To avoid that one could require that the plateau between $\phi=0$ and $\phi_{\max}$ is long enough to generate successful inflation. This requirement is fulfilled for all $|\lambda_1|,\lambda_2 \sim \mathcal{O}(1)$. Nevertheless for $|\lambda_1|,\lambda_2\gg 1$ one can obtain a short plateau for which the last 60 e-folds of inflation are partially generated by the inflaton in the $\phi>\phi_{\max}$ regime. At the end the field rolls down to $\phi=0$, oscillates around the minimum and reheats the universe. 
\\*

Let us assume that the field is somehow bigger than $\phi_{\min}$ and the $\lambda_2$ term dominates the potential. Then $V\simeq \lambda_2 \exp(-2\sqrt{2/3}\,\phi) U_{S}^3\propto \exp(4\sqrt{2/3}\,\phi)$. Such a potential cannot generate the slow-roll evolution of the field since the slow-roll parameter $\epsilon = 16/3 \gg1$. If the field cannot escape the minimum at $\phi_{\min}$ it oscillates around it and due to the existence of the cosmic friction finishes its evolution at $\phi=\phi_{\min}$. If $V(\phi_{\min})>0$ then one obtains an exact de-Sitter solution for the evolution of the space-time. This kind of inflation suffers from lack of consistency with PLANCK \cite{Ade:2013zuv} and BICEP \cite{Ade:2014gua} data because of the perfectly flat power spectrum of primordial inhomogeneities. Another problem is a lack of the graceful exit, which ends inflation and allows to reheat the universe after inflation. A possible solution to that problem may be the quantum tunnelling, which will be discussed in the following parts of this paper. The possible solution to the flat power spectrum problem could be to introduce the additional curvaton field responsible for the generation of the primordial inhomogeneities. The potential of the curvaton should not be perfectly flat in order to generate perturbations consistent with observational data. Thus, the de Sitter inflation should not last for too long. Otherwise the curvaton would vanish before the tunnelling. Another issue appears when $\phi_{\min}$ is a global minimum of $V$. Then, the field which managed to reach $\phi_0$ may tunnel to $\phi_{\min}$, which changes the effective value of gravitational constant. In such a case the theory would not recover GR at low energies.
\\*

\begin{figure}[h]
\centering
\includegraphics[height=5.2cm]{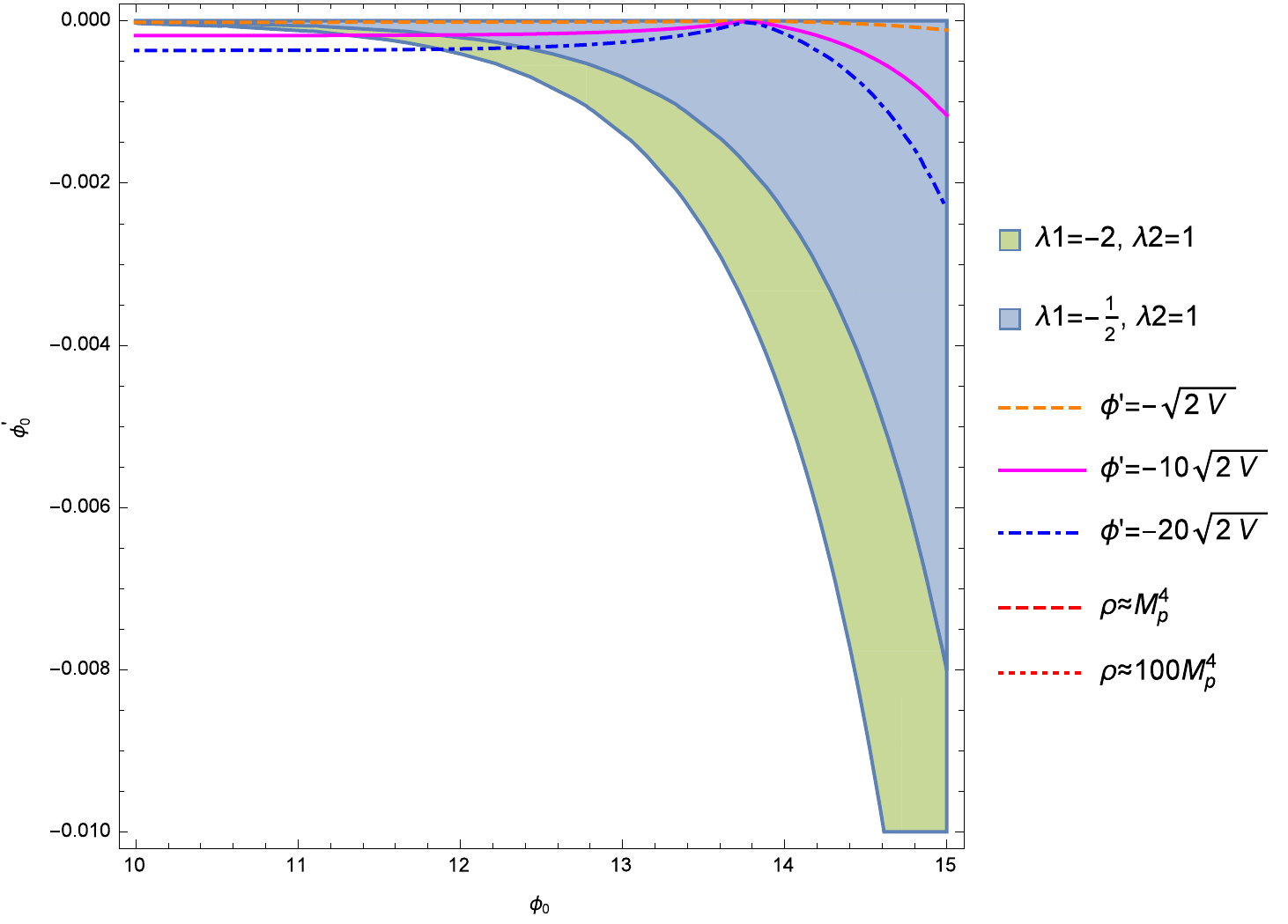}
\hspace{0.5cm}
\includegraphics[height=5.2cm]{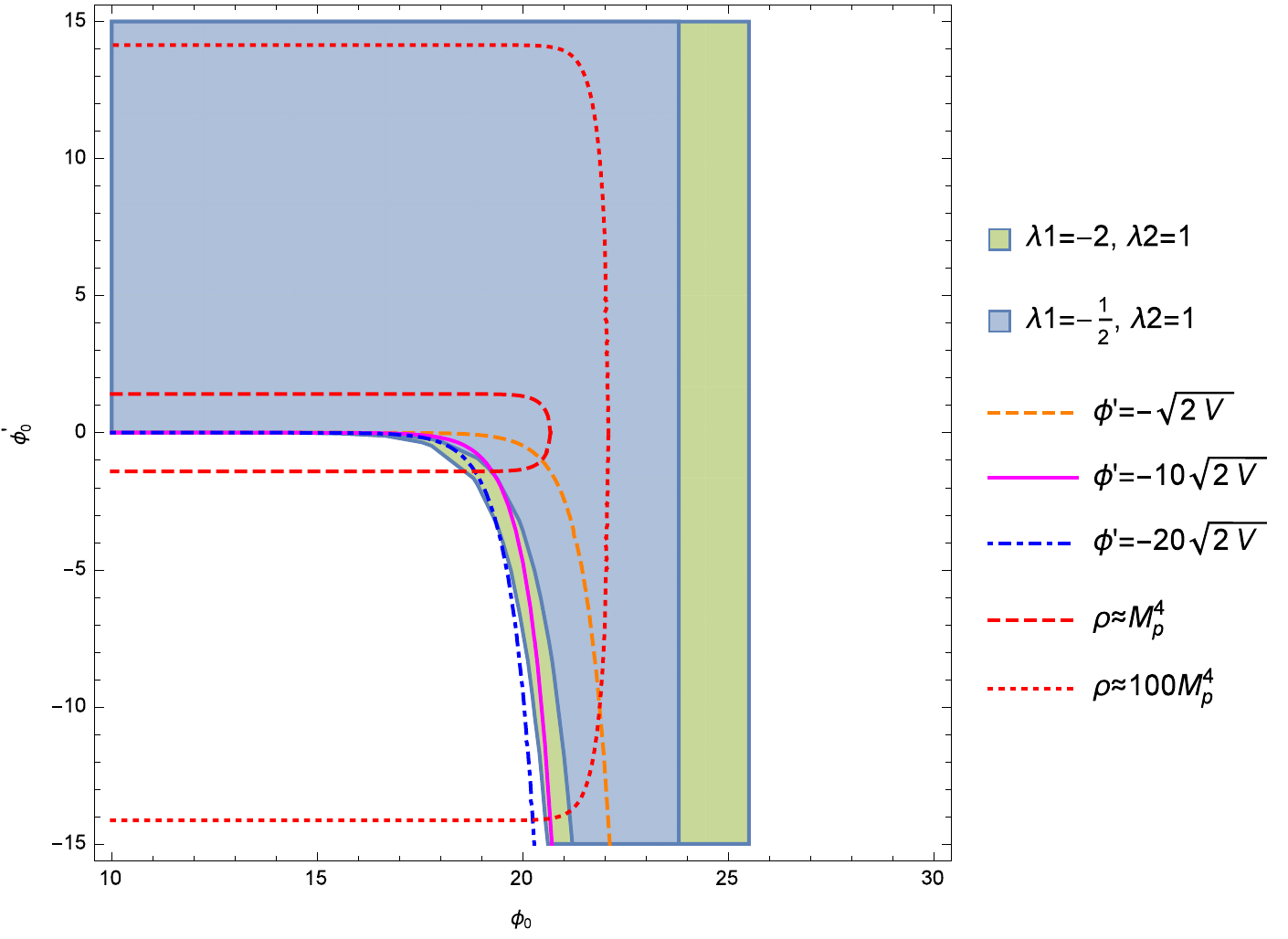}
\caption{\it Both panels present regions in the phase space of initial conditions, for which the GR minimum cannot be reached. Red curves correspond to $\rho_0 = M_{p}^4$ and $\rho_0 = 100 M_{p}^4$ (dashed and dotted red lines respectively). As shown in the right panel the field can reach the GR vacuum only for $\phi_0'^2 \gg V(\phi_0)$ and only for $\phi_0' <0$. Even super-Planckian values of $\phi_0'>0$ would give the evolution which ends at $\phi_{\min}$.} 
\label{fig:initialconditions1}
\end{figure}

Fig.~\ref{fig:initialconditions1} shows the areas in the phase space of initial conditions that lead to $\phi\to\phi_{\min}$ at the late times. The region with enables $\phi$ to reach $\phi=0$ lies in the $\phi_0'^{2}\gg V(\phi_0)$ regime, so the kinetic term domination is required to reach the GR minimum via the classical evolution of the field. Let us note that no positive value of $\phi'_0$ allows $\phi$ to reach the GR minimum.

\subsection*{Planck scale limit}

One can ask when the initial conditions give the potential energy of order of the Planck scale. For given values of $\lambda_1$ and $\lambda_2$ let us define $\phi_{pl}$ such as $V(\phi_{pl})=M_{p}^4$. Whenever $\phi>\phi_{pl}$ the potential term is bigger than the Planck scale and quantum gravity corrections should be included. The potential $V(\phi)$ reaches the Planck limit when the $\lambda_2$ term dominates. Thus, one can easily assume that in that limit $U\simeq \lambda_2 U_{S}^3$. Then, from the definition of $\phi_{pl}$ one finds
\begin{equation}
\phi_{pl} = \sqrt{\frac{3}{2}}\log\left(\frac{2 \sqrt{2} \lambda _2^{1/4}}{3^{3/4}M^{3/2}}\right) \, .
\end{equation}
The biggest contribution to $\phi_{pl}$ comes from $M$. On the other hand the $\lambda_2$ dependence is very weak. For all realistic values of $\lambda_2$ and for $M=1.5\times 10^{-5}$ one finds $\phi_{pl}\simeq 20$. 
\\*

The Planck scale as the limit of the classical theory is not set precisely. Different models of quantum gravity predict different scales of validity of the classical theory. For instance in loop quantum cosmology (LQC) \cite{Bojowald:2006da}, the maximal energy density of the system, at which the quantum corrections are the most significant is of order of $250M_{p}^4$. This would allow bigger values of $\phi$, which lie within the classical limit of the theory. The situation becomes more complicated when one considers LQC corrections to modified theories of gravity \cite{Artymowski:2013qua,Zhang:2012em}. Then, the maximal energy scale depends on the frame in which the system is quantized, and on the value of $\phi$ at the moment of maximized energy density. Thus, one cannot set one scale that would clearly limit the $\phi$ from the point of view of the validity of the classical theory. The $\phi_{pl}$ is just a realistic approximation, especially since the slope is very steep and any increase of $\phi$ over $\phi_{pl}$ increases the energy density of the system by orders of magnitude. The issue of the LQC corrections to this model shall be a future extension of the analysis performed in this paper.

\subsection*{The problem of initial conditions}

Let us note that due to stronger limits on the tensor-to-scalar ratio $r$ the PLANCK data favour the plateau-like potentials (like in Starobinsky inflation) and disfavour the power-law potentials (like $m^2\phi^2$). Nevertheless, the plateau-like models of inflation struggle with some difficulties mentioned in the Ref. \cite{Ijjas:2013vea}. One of the problems, which appears within the analysis of the Starobinsky model, is the fact that the Einstein frame potential is limited from above by the scale of the order of $M^2\ll M_{p}^4$. Thus, if one would set initial conditions at the Planck scale the potential term would be always subdominant and the $(\partial_i \phi)^2$ term may dominate the universe, which would lead to strong inhomogeneities. While potentials without the upper bound generate the Planck scale initial conditions $(\partial_i \phi)^2 \sim \dot{\phi}^2 \sim V(\phi)$ \footnote{For argument see e.g. the Ref. \cite{Kofman:2002cj}.}, for which the potential term can dominate the evolution of the field and homogenize the universe.  
\\*

For the Starobinsky potential with higher order corrections this problem seems to be weakened. The steep, exponential slope at big $\phi$ reaches the Planck scale around $\phi\sim 20$ and the potential can have significant contribution to the initial energy density. Nevertheless the problem of initial conditions is still an issue of our model, since inflation cannot start on the slope and inhomogeneities can still dominate the evolution of the universe when the field is on its way to the Starobinsky plateau. The other issue appears for $\lambda_1<-\lambda_s$. In order to pass local minimum and reach the plateau we require the kinetic term domination at the Planck scale. Thus, before reaching the beginning of inflation the inhomogeneous terms may become significant and spoil our analysis based on FRW metric.
\\*

Let us note that the slight modification of the Starobinsky potential based on the $f(R) = R + \alpha R^{n}$, where $n\lesssim 2$ is even better cure for the problem of initial conditions. For $n\neq2$ the Einstein frame potential is of the form  \cite{Artymowski:2014gea,Artymowski:2014nva}
\begin{equation}
V \propto V_S\times \left(\exp\left(\sqrt{\frac{2}{3}}\phi\right)-1\right)^{\frac{2-n}{n-1}}\quad \underrightarrow{\phi\gg 1}\quad \exp\left(\frac{2-n}{n-1}\sqrt{\frac{2}{3}}\phi\right)\, ,
\end{equation}
where $V_S:=U_S/\varphi^2$. This potential is not limited from above, so $V(\phi_{pl})$ can have large contribution to the energy density at the Planck scale. Another advantage of this model is that inflation can start for any $\phi$ and therefore to obtain successful inflations without significant inhomogeneities one needs only one bubble at the Planck scale in which $V(\phi)$ dominates.


\section{Inflationary dynamics} \label{sec:infl2}

\subsection*{Inflation with and beyond the slow-roll approximation}

In this section we will discuss the slow-roll approximation and inflation in the Einstein frame. Inflation takes place on the plateau, where the influence of the $\lambda_2$ term is very small, but as we will show there are also other mechanisms generating inflation, which 
strongly rely on $\lambda_2$.
We want to investigate how non-zero values of $\lambda$ parameters deviate the inflation away from the Starobinsky model. Let us assume that $\phi''\ll V_\phi$. Then, for $\rho_M = P_M = 0$ one obtains
\begin{equation}
3\mathcal{H}\phi' + V_\phi \simeq 0\, , \qquad 3\mathcal{H}^2 \simeq V \, .\label{eq:SReoms}
\end{equation}
This approximation holds for non-zero values of $V_\phi$, so one cannot use Eq. (\ref{eq:SReoms}) at $\phi = \phi_{\max}$. Thus we will use the slow-roll equations for $\phi\neq\phi_{\max}$. The cosmic inflation takes place as long as following slow-roll parameters are much smaller than one
\begin{equation}
\epsilon: = \frac{1}{2}\left(\frac{V_\phi}{V}\right)^2  \, , \qquad \eta := \frac{V_{\phi\phi}}{V} \ . \label{eq:SRparameters}
\end{equation}
The number of e-folds generated during the inflation is in the slow-roll approximation equal to 
\begin{equation}
N = \int_{t_i}^{t_f} \mathcal{H}d\tilde{t} \simeq \int^{\phi_i}_{\phi_f} \frac{V}{V_\phi}d\phi \, , \label{eq:Efolds}
\end{equation}
where indexes $i$ and $f$ refer to initial and final moments of inflation respectively. Namely, $t_i$ is the first moment when both slow-roll parameters are smaller than one and $t_f$ is the moment when any of slow-roll parameters becomes bigger than one. $\phi_f$ is independent of higher order correction to the potential, but in general $\phi_i$ may depend on $\lambda_1$, $\lambda_2$ and initial conditions chosen for the evolution. 
\\*

For $\lambda_1< - \lambda_s,\ \lambda_2>0$ there is a local maximum at $\phi_{\max}$, which may influence {markedly}  the total number of e-folds generated during inflation. One can fine-tune initial conditions to fit the following evolution of the inflaton: $\phi$ has just enough energy to climb to the $\phi_{\max}$, so at the local maximum its kinetic energy is almost zero. Then $\phi$ stays very long around $\phi_max$, which in principle may generate unlimited number of e-folds according to the classical evolution of $\phi$. Even if $\lambda_1,\ \lambda_2$ are so big that the Starobinsky plateau is too short to generate at least $60$ e-folds of inflation, one could generate sufficiently large $N$ during the evolution around $\phi_{\max}$. The infinite production of e-folds appears for fine-tuned initial conditions which lead to $\phi = \phi_{\max}$ and $\phi' = 0$. 
\\*

This effect is limited by the existence of quantum fluctuations of the inflaton. The vev of the inflaton evolves due to the classical evolution and due to quantum fluctuations. During one Hubble time those effects modify the value of the field by $\delta\phi_C \sim \phi'/\mathcal{H}$ and $\delta\phi_Q \sim \mathcal{H}/2\pi$ respectively. Whenever $\delta\phi_Q \gtrsim \delta\phi_C$ the classical equations of motion cannot be used to describe the evolution of $\phi$. In our case this means that whenever the field rolls up or down on the attractor trajectory around $\phi_{\max}$ and satisfies $\delta\phi_Q \gtrsim \delta\phi_C$ its vev is pushed after one Hubble time towards $\phi_{\min}$ or GR vacuum by quantum fluctuations of the inflaton. To see how {efficient} quantum fluctuations are let us define the region on the attractor which gives $\delta\phi_Q\geq\delta\phi_C$ by $\phi\in(\phi_{\max}-\Delta\phi,\phi_{\max} + \Delta\phi)$, where $\Delta \phi$ satisfies condition $\delta\phi_Q(\phi_{\max}\pm\Delta\phi) = \delta\phi_C(\phi_{\max}\pm\Delta\phi)$. Since the Eq. (\ref{eq:SReoms}) is satisfied at the attractor one finds
\begin{equation}
\delta\phi_Q \simeq \frac{1}{2\pi}\sqrt{\frac{V}{3}} \, , \qquad \delta\phi_C \simeq \frac{V_\phi}{V} \quad \Rightarrow \quad \Delta\phi\simeq \frac{1}{8\pi}(-M/6\lambda_1)^{1/3} \, .
\end{equation}
For $\lambda_1 = -1$, $M = 1.5\times 10^{-5}$ this gives $\Delta\phi\simeq 5.4\times 10^{-4}$, which is two orders of magnitude bigger than the typical value of $\delta\phi_Q$ for this particular value of M. During one Hubble time from our comoving Hubble radius one obtains $e^3$ causally independent regions. In some of them $\delta\phi_Q<0$ and the field is pushed toward the Starobinsky plateau. Nevertheless one would need $100$ such a fluctuations (or few with extraordinary amplitude) to leave the part of the attractor with quantum fluctuations domination. It is indeed highly improbable event and in most of horizons generated while $|\phi-\phi_{\max}|<\Delta\phi$ it shall never happen. Thus, whenever the field reaches the attractor at $|\phi-\phi_{\max}|<\Delta\phi$ the field could never reach the classical evolution domination regime. Thus one obtains the eternal inflation. To avoid this problem one would need $\lambda_1> 3\times 10^{-4}M^{-2}\sim 10^6$, which gives $\Delta\phi<\delta\phi_Q$. In such a case $\phi$ would leave the regime of quantum fluctuations domination after $t\sim 1/H$. The other issue is that for $|\lambda_1|,\lambda_2\to 0$ one finds that $\Delta\phi$ and $\phi_{\max}$ are growing like $\lambda_1^{-1/3}$ and $-\log{\lambda_1}$ respectively. Since power-law functions are growing faster that the logarithm one expects the whole plateau to be inside the $(\phi_{\max}-\Delta\phi,\phi_{\max}+\Delta\phi)$ region. Fortunately this behaviour of $\Delta\phi$ does not hold for $|\lambda_1|,\lambda_2 \ll 1$ and $\Delta\phi$ region is just swallowed by the other region of eternal inflation at $\phi>17.3$. In principle the quantum fluctuations around the maximum could lead to the so-called topological inflation. This issue was partially analysed in the Ref. \cite{Kamada:2014gma} and will be extended in our further work.
\\*

\begin{figure}[h]
\centering
\includegraphics[height=5cm]{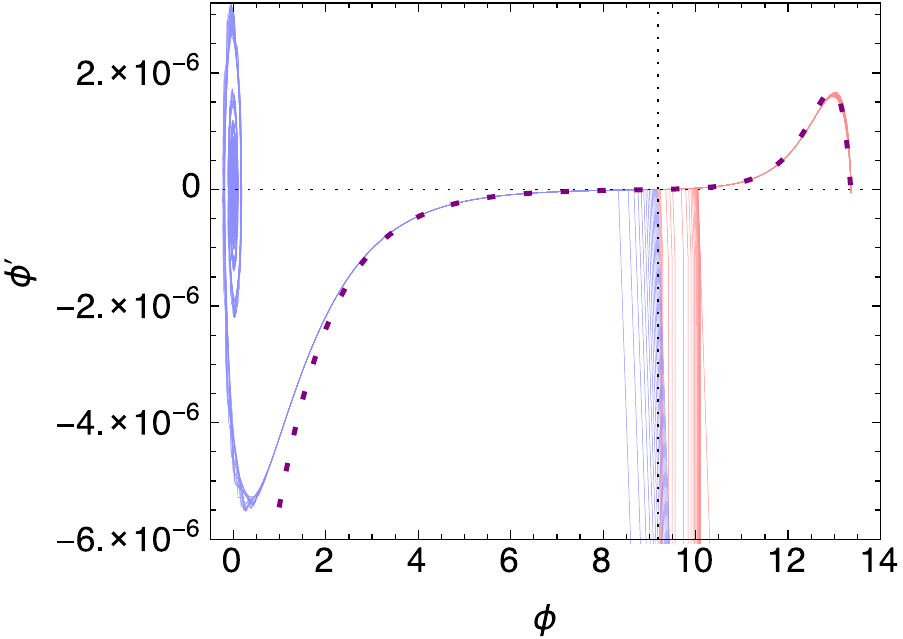}
\hspace{0.5cm}
\includegraphics[height=5cm]{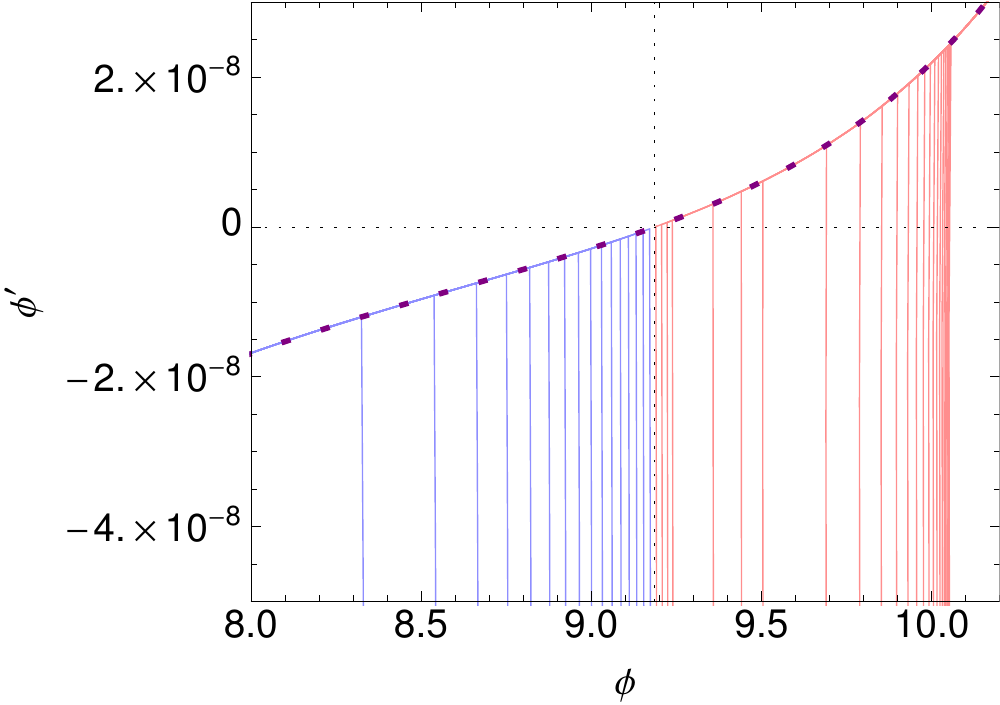}
\caption{\it Both panels present the phase space analysis for $\lambda_1 = -\lambda_2 = -1$. Different initial conditions give the evolution, which leads to one of two regions of the attractor: a) The blue one gives inflationary trajectory at the Starobinsky plateau and oscillations around $\phi = 0$. b) The red one leads to $\phi_{\min}$ and the De Sitter solution. Dashed purple line represents the slow-roll approximation, which describes the evolution the attractor from $\phi \sim 1$ to $\phi_{\min}$. Axes originates at $(\phi_{\max},0)$.} 
\label{fig:attractor}
\end{figure}

As shown at the Fig. \ref{fig:attractor} the inflaton during its evolution reaches the attractor  separated into  two regions. The one  that exists for $\phi<\phi_{\max}$ leads to Starobinsky-like inflation and to oscillations around the GR vacuum. Initial conditions, for which $\phi$ reaches the attractor at $\phi>\phi_{\max}$ give the evolution which always ends at $\phi = \phi_{\min}$ and $\phi' = 0$. Those regions of the attractor are separated by the region where quantum fluctuations dominate around $\phi_{\max}$, where classical equations of motion do not hold.

\begin{figure}[h]
\centering
\includegraphics[height=5cm]{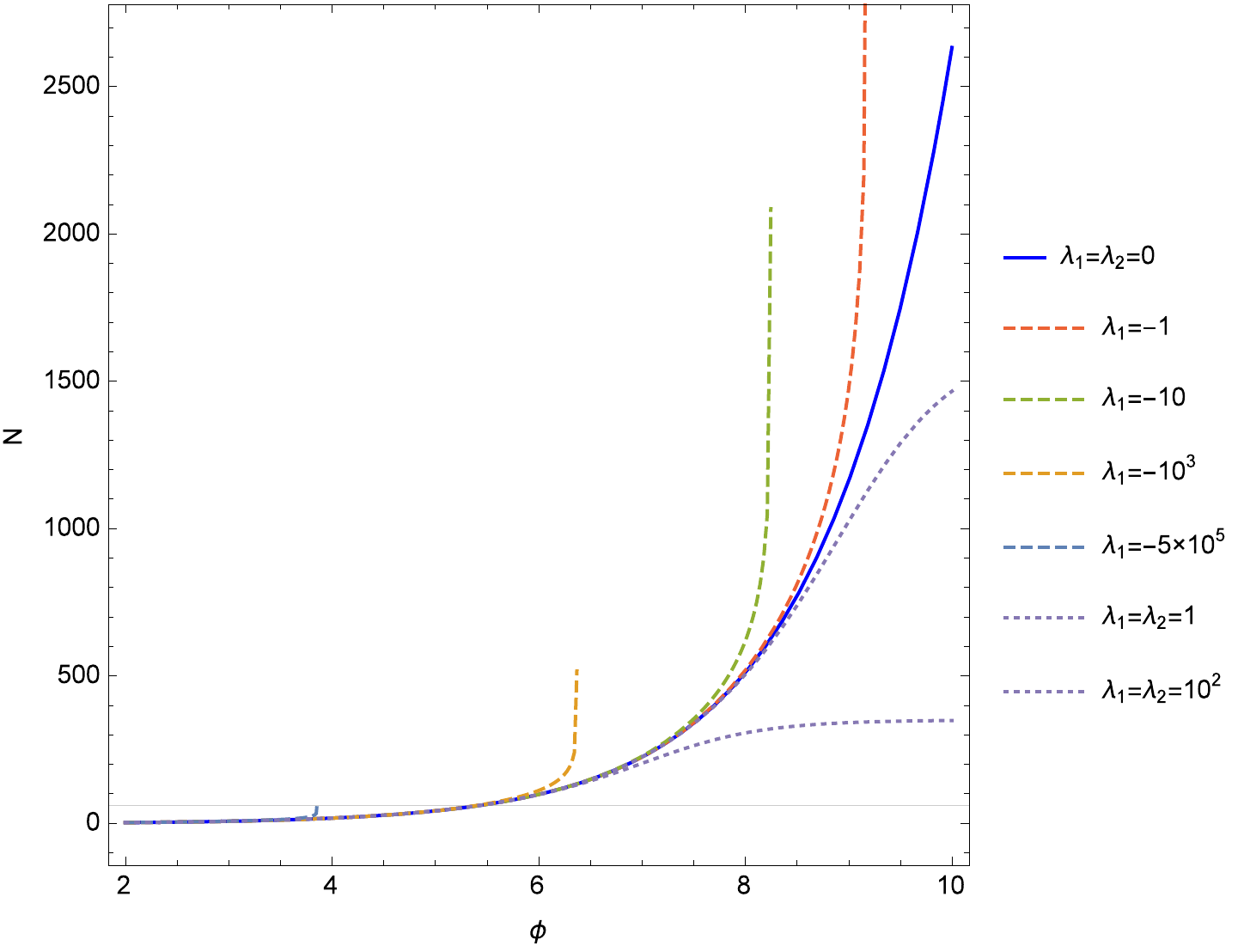}
\hspace{0.5cm}
\includegraphics[height=5cm]{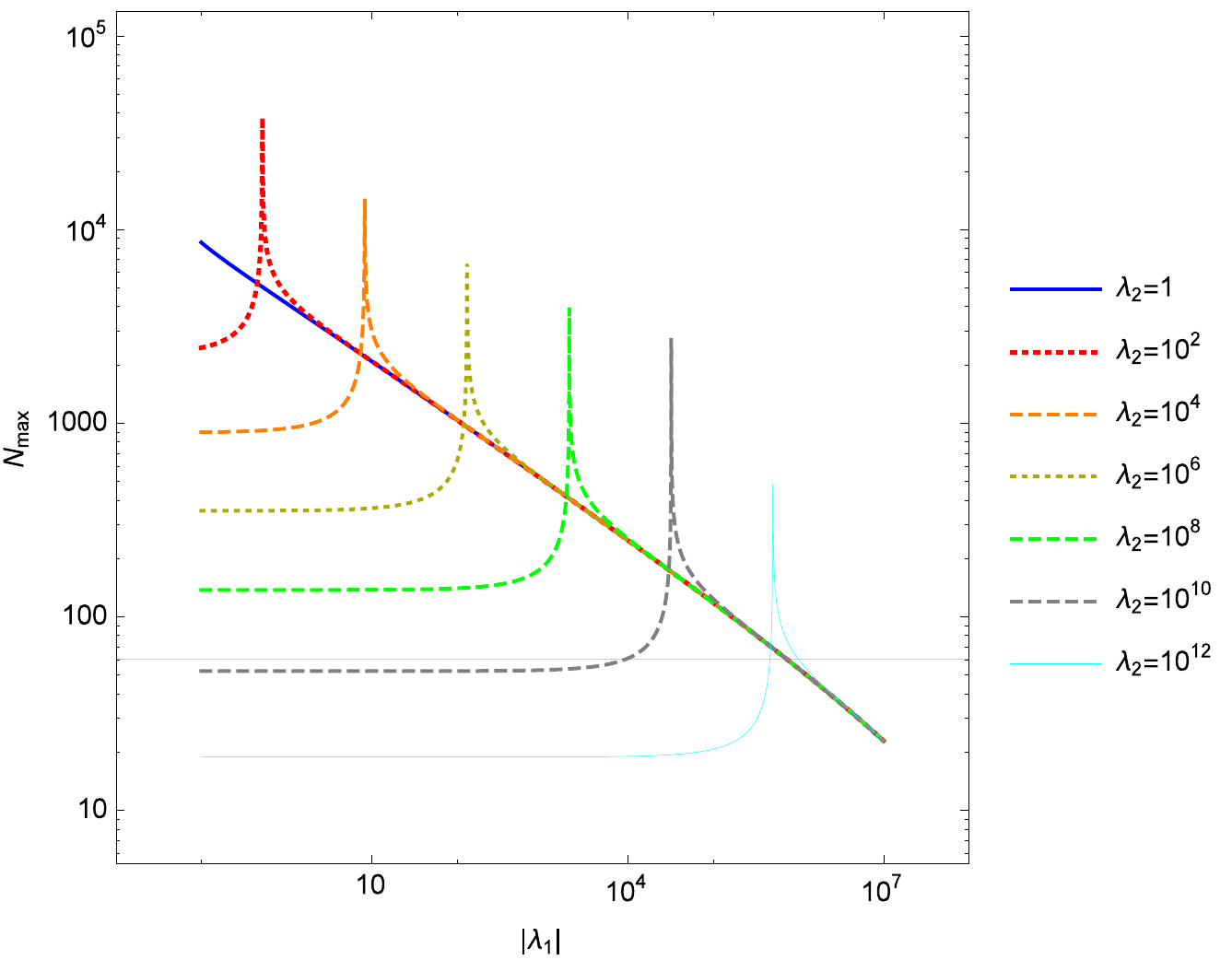}
\caption{\it Left panel: Number of e-folds generated during inflation as a function of initial field value. The $\lambda_1=\lambda_2=0$ corresponds to Strobinsky, and other lines illustrate how fast is the departure from this simplest case due to corrections. The value of $\lambda_2$ in all cases with negative $\lambda_1$ was fixed be requiring that the value of the potential in the minimum is $M^2/10$.\hspace{\textwidth}
Right panel: $N_{\max}$ for different values of $\lambda_1$ and $\lambda_2$, where $\lambda_1<0$. Peaks correspond to the saddle point of the potential at $\lambda_1 = -\lambda_s$. Flat lines left from peaks: minimum and maximum do not exist and inflation may start as soon as the slope ends on the plateau. The $\lambda_2 = 1$ line right from peaks: inflation may start as soon as $\phi<\phi_{\max} - \Delta\phi$. {All the lines converge to the same function for large enough $\lambda_1$ because after the new minimum appears number of e-folds to the leading order is only a function of $\lambda_1$}}
\label{fig:Nmax}
\end{figure}

As mentioned previously in this section whenever the inflaton reaches the attractor at $|\phi-\phi_{\max}| \leq \Delta\phi$ one obtains the quantum fluctuations domination regime and eternal inflation. To avoid that let us assume that $\phi_i<\phi_{\max} - \Delta\phi$. If $\phi_i < \log \left(-M^{-2}\lambda _1^{-1}\right)/\sqrt{6}$ then one can expand the number of e-folds in terms of the mass M. Then, for $\phi_i = \phi$ one obtains
\begin{equation}
N \simeq \frac{1}{4} \left(3 e^{\sqrt{\frac{2}{3}} \phi } - \sqrt{6} \phi -\frac{9}{16} e^{4 \sqrt{\frac{2}{3}} \phi } M^2 \text{$\lambda $1}\right) \, .
\end{equation}
In fact the biggest amount of e-folds can be generated when $\phi$ is closer to $\phi_{\max}-\Delta\phi$ since in this regime $V_{\phi}$ is  smallest. In such a case the linear approximation collapses and one need to employ numerical methods. Keeping that in mind let us define the maximal amount of e-folds generated during the classical evolution of the field by 
\begin{equation}
N_{\max} = \int^{\phi_{\max} - \Delta\phi}_{\phi_f}\frac{V}{V_\phi}d\phi \, .
\end{equation}
Its values for different $\lambda_1$ are presented at the Fig. \ref{fig:Nmax}. We require that $N_{\max} > 60$, which gives $\lambda_1\lesssim 5 \times10^5$. In fact $\lambda_1$ could be much bigger but this would require $\lambda_1 = -\lambda_s$ and a saddle point inflation. One can see that for $\lambda_1<-\lambda_s$ (right from the peak for given $\lambda_2$ at the Fig. \ref{fig:Nmax}) the $N_{\max}$ does not depend on $\lambda_2$. This comes from the fact that $\phi_{\max}$ and $\Delta\phi$, which determinate maximal allowed $\phi_i$, depend mostly on $\lambda_1$ and very weekly on $\lambda_2$. 
\\*

The primordial inhomogeneities generated in this model do not deviate much from the Starobinsky inflation as long as $\phi_{\max}$ is separated from the scale of freeze-out of perturbations by few $M_{p}$. In such a case the last 60 e-folds of inflation happen in the regime of negligible influence of higher order terms. On the other hand, for $\phi_{\max}\sim 5$ or for the saddle point inflation one expects significant changes in the shape of power spectrum and in the value of $M$ as compared to the Starobinsky model. This issue is currently under investigation and it shall be discussed  in a forthcoming paper.

\subsection*{Eternal inflation for large $\phi$}

As mentioned above the evolution of $\phi$ may, under some circumstances, be strongly influenced by quantum fluctuations. For the Starobinsky inflation one finds that 
$$\delta\phi_Q > \delta\phi_C \qquad\text{for}\qquad \phi>\phi_c = -\sqrt{\frac{3}{2}}\log\left(\frac{\sqrt{6}M}{16\pi}\right) \, .$$
For $M = 1.5\times 10^{-5}$ one finds $\phi_c \simeq 17.3$. For the Starobinsky inflation with $\phi_0>\phi_c$ one obtains the uplifting of $\phi$ by quantum fluctuations, which disables the field to roll down. In such a case inflation never ends, which is the so-called eternal inflation. As mentioned in \cite{Broy:2014sia} the higher order terms in the potential may solve this problem. Indeed, if any of higher order terms would generate the slope or the minimum at some $\phi<\phi_c$ one would be free from the eternal inflation. To satisfy this condition one needs
\begin{equation}
\lambda_1 > \frac{4}{3M^2}e^{-2\sqrt{\frac{2}{3}}\phi_c} = \frac{1}{32\pi^2} \simeq 3.2 \times 10^{-3}\, , \quad \lambda_2 >\frac{16}{9M^4}e^{-4\sqrt{\frac{2}{3}}\phi_c} = 2^{-10}\pi^{-4}  \simeq  2.4 \times 10^{-4} \, .
\end{equation}
This result does not depend on $M$.


\section{Thermal corrections to the evolution of the system} \label{sec:Temp}

\subsection*{Introduction}

When the field is trapped in $\phi_{\min}$ with $V(\phi_{\min}) = 0$ one can try to evolve it to $\phi=0$ with the help of the thermal corrections. The non-zero temperature of the universe may be obtained by the dissipation of the inflaton's energy density into relativistic degrees of freedom from the standard model (SM). The idea of thermal corrections to the potential has been developed in the context of the ``Mexican hat'' potential  \cite{Linde:2005ht,Bezrukov:2014ipa}, where structure of the vacuum breaks spontaneously the symmetry of potential. While the temperature it growing the potential is modified due to the existence of the effective potential term proportional to $T^2\phi^2$. This additional term fills the minimum, pushes the field towards zero veg and restores the symmetry of the potential. 
\\*

In our case there may be two sources of non-zero temperature in the inflationary Universe before the reheating. First of all one can reasonably assume that the universe before inflation was warm or that the inflaton has decayed into relativistic degrees of freedom. The decay of the inflaton at $\phi=\phi_{\min}$ is in principle the same as the post-inflationary reheating of the universe. The $\phi$ is coupled to scalars, fermions and vectors of the SM (or to other particles, which can decay into the SM degrees of freedom), which are being produced during the oscillations of the field around $\phi_{\min}$. If $V(\phi_{\min})>0$ then the vacuum energy generates the de Sitter space-time, in which the energy density of relativistic particles will decrease like $e^{-4Ht}$. Thus, the universe would remain cold and thermal corrections will not modify the potential.
\\*

Let us define the Einstein frame potential with non-zero temperature by $\mathcal{V}(\phi,T) = V(\phi) + \Delta V(\phi,T)$, where $\Delta V(\phi,T)$ is the thermal correction to the potential. The thermal correction to the Einstein frame potential has a following form \cite{Kapusta:2006pm}
\begin{equation}
\Delta V = \frac{T^2}{24}V_{\phi\phi}\, ,
\end{equation}
where $T$ is the Einstein frame temperature. This means, that the Einstein frame radiation energy density is equal to $\rho_R = \pi^2/30\  g_\star T^4 $, where $g_\star$ is the number of relativistic degrees of freedom. In general $g_\star = g_\star (T)$, since $g_\star$ may decrease with the energy scale. 

\subsection*{The features of the Einstein frame potential with the thermal correction}

\begin{figure}[h]
\centering
\includegraphics[height=5.8cm]{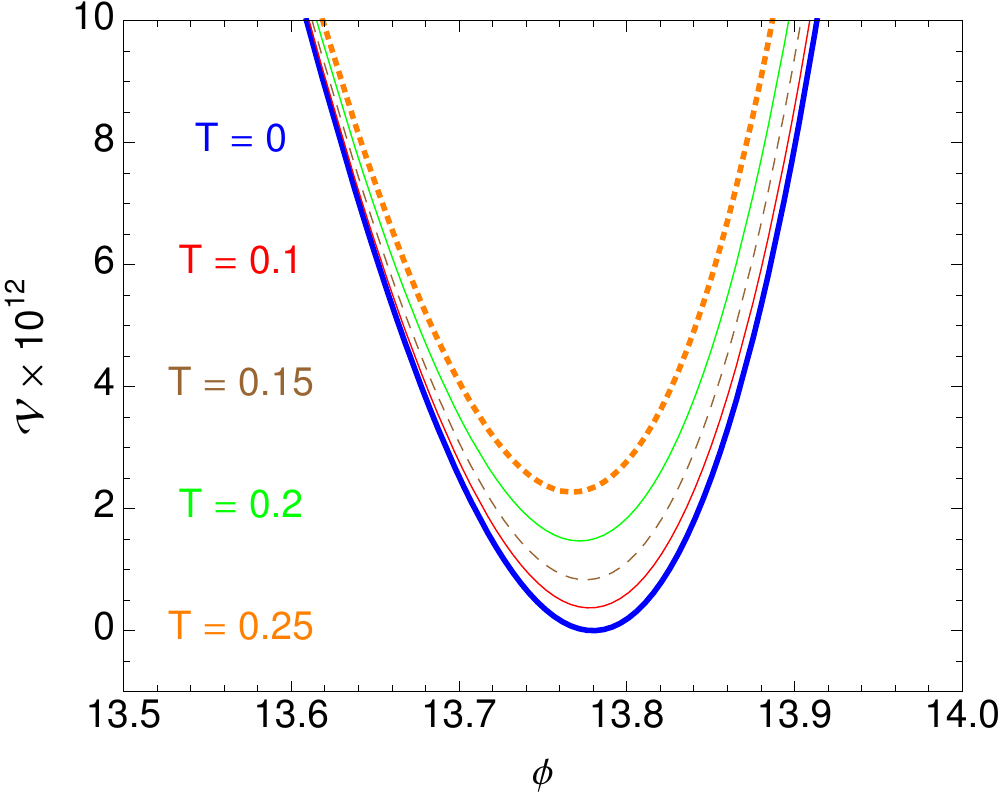}
\hspace{0.5cm}
\includegraphics[height=5.8cm]{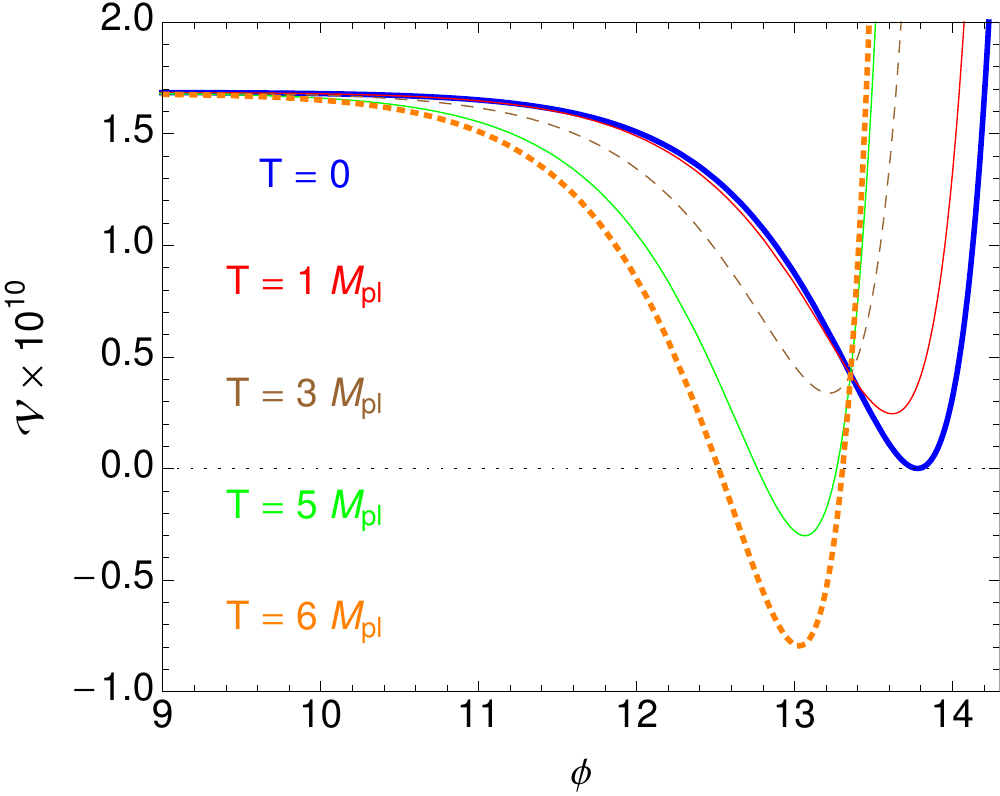}
\caption{\it Both panels present $\mathcal{V}(\phi,T)$ for $\lambda_1 = -1$, $\lambda_2=1$ and for different values of $T$. Left Panel: maximal realistic values of $T$. The minimum is being uplifted, but the inflaton still cannot escape to the GR vacuum. Right panel: even for unrealistic values of $T$ the minimum will not be filled by the thermal correction. For $T>3M_{p}$ the $\mathcal{V}_{\min}$ starts to decrease towards the negative values. These results remain the same even for $|\lambda_1|,\, \lambda_2 \gg 1$.} 
\label{fig:VT}
\end{figure}

In this section we will apply the procedure introduced in the Sec. \ref{sec:EinAn} into the $\mathcal{V}(\phi,T)$ case. As previously, we will use the fact that $\exp(\sqrt{2/3}\, \phi_{\min})\sim M^{-1}$ and we will assume that $|\lambda_1|,\lambda_2 < M^{-1}$. Since $\rho_R\ll M_{p}^4$ we also assume that $T\ll 1$. Then $\mathcal{V}$ has a minimum at $\phi^T_{\min}$ defined by
\begin{equation}
\phi^T_{\min} \simeq \frac{1}{2}\sqrt{\frac{3}{2}}\left(\log \left(-\frac{2 \lambda _1}{3\lambda _2\,M^2} \right)-\frac{T^2}{3}\right) = \phi_{\min} - \frac{1}{6}\sqrt{\frac{3}{2}}T^2\, .\label{eq:phiMinT}
\end{equation}
One can see that the correction is rather small and it does not significantly shifts the minimum. The potential energy density at $\phi = \phi^T_{\min}$ is following
\begin{equation}
\mathcal{V}_{\min} \simeq \frac{3M^2 }{16 \lambda _2}\left(4\lambda _2-\lambda _1^2\right)+\frac{M^2 \lambda _1^2}{24 \lambda _2}T^2 = V_{\min } + \frac{M^2 \lambda _1^2}{24 \lambda _2}T^2 \, .\label{eq:VTmin}
\end{equation}
The non-zero temperature is uplifting the potential in the minim. However, this effect is way to small to fill the minimum and to push $\phi$ towards the GR vacuum. Even if one would lift $T$ towards the Planck scale the minimum at $\phi^T_{\min}$ wouldn't be filled. Higher order corrections in $T$ would enter the Eq. (\ref{eq:VTmin}) and the energy density would start to decrease with temperature. This issue is depicted in Fig. \ref{fig:VT}.
\\*

The thermal corrections could be also treated as a way to make the minimum at $\phi_{\min}$ more shallow while the field is on the slope. In such a case the region in the phase space of initial conditions which allowed $\phi$ to reach the GR vacuum would be bigger. Unfortunately, due to the fact that $\Delta V\ll V$ for $T\ll M_{p}$, this effect would be very small. Thus, the non-zero temperature of the universe shall not significantly change the evolution of the system.


\section{Quantum tunnelling from meta-stable minima}\label{sec:Tunnel}

Whenever a new minimum appears in the potential due to higher order corrections {(i.e. for $\lambda_1<-\lambda_s$)}, the field may be unable to cross the barrier due to classical evolution if its initial speed is to small. In this section we will discuss whether it is possible for the field to cross the barrier due to quantum fluctuations.  

We will focus mainly on two limiting cases for such an event. First one is the Hawking-Moss effect (HM) \cite{Hawking:1981fz} describing a simple fluctuation of the field, large enough for it to pass the barrier. The second case is quantum tunnelling. Here we will use the standard formalism of Coleman and De Luccia (CDL) \cite{Coleman:1980aw}, assuming that vacuum decay proceeds through nucleation of true vacuum bubbles within our false vacuum. Details of this calculation are described in the following subsection\ref{sec:CDLinstantons}. During this analysis we show two examples of potentials with fixed values of the false vacuum minimum $V_{\min}=M^2/10$ and $V_{\min}=M^2/100$. In the remaining part of this section we use these conditions and calculate $\lambda_2$ for each value of $\lambda_1$. While we use a precise numerical result, the analytical estimate coming from \eqref{eq:Vmin} will be sufficient to understand these results, namely
\begin{equation}
\lambda_2 \approx \frac{\lambda_1^2}{4- \frac{16}{3M^2}V_{\min}}.
\end{equation} 

\begin{figure}[h]
\centering
\includegraphics[height=5.8cm]{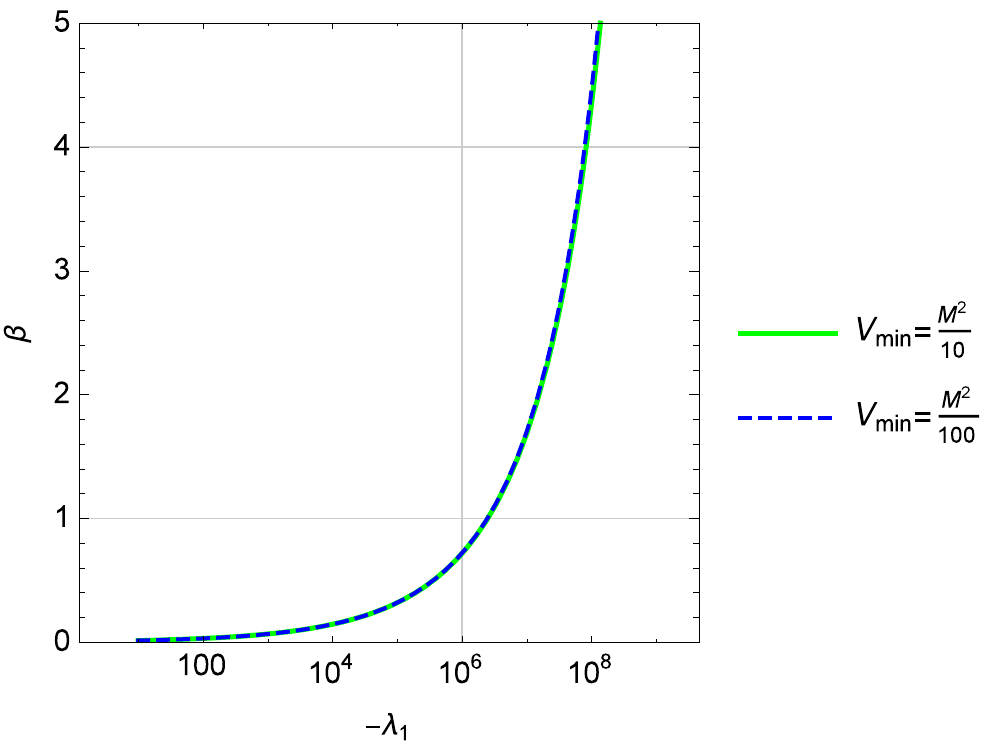}
\caption{\it Parameter $\beta$ from \eqref{eq:beta} in our two examples. The CDL solutions exist only if $\beta>4$. And below $\beta=1$, only HM solutions exist. The value of $|\lambda_1|=10^6$ is the largest allowing 60 e-folds of inflation on the {plateau. For such a big $\lambda_1$ the topological inflation at $\phi = \phi_{\max}$ is negligible, since $\Delta\phi<H/2\pi$.}}  
\label{fig:betaplot}
\end{figure}

Existence of the plateau between the two vacua is crucial in vacuum stability analysis because CDL instantons cannot penetrate a barrier which is too flat. The crucial parameter here is \cite{Hackworth:2004xb}
\begin{equation}\label{eq:beta}
\beta=3 \frac{V_{\phi\phi}(\phi_{\max})}{V_{\max}}\, ,
\end{equation} 
for $\beta > 4$ both CDL and HM solutions exist. We can immediately see a problem because this parameter is proportional to $\eta$  slow-roll parameter from \eqref{eq:SRparameters}. This suggests we can either have CDL solutions or large number of e-folds during inflation, not both. This suspicion is confirmed by Figure~\ref{fig:betaplot} which shows parameter $\beta$ for our two example scenarios. 

The barrier penetration through CDL instantons becomes possible only for very large values of parameters $\lambda$. This happens because the larger the couplings are the shorter and less flat the potential barrier becomes. One can see that when $\beta$ reaches $4$ these potentials are already excluded by lack of suitable inflation as discussed in Section~\ref{sec:infl2}. {The only way to generate over $60$ e-folds for such a big $\lambda$ parameters would be to assume that $\lambda_1 = -\lambda_s$. However. such a potential would not have a minimum and thus the issue of quantum tunnelling would not exist}.
\\*

{For both type of instantons considered in this paper one needs to calculate the decay probability of the metastable vacuum defined by}
\begin{equation}\label{eq:tau}
\Gamma =  A e^{-S},
\end{equation} 
{where $S$ is an action of the considered type of instantons.} The prefactor A is derived from quantum corrections to the bounce and always is less important than the exponent. Action of an HM instanton is simply the difference between the action of homogeneous solution for field in the false vacuum and on top of  the barrier at $\phi_{\max}$
\begin{equation}\label{eq:tau}
S_{\textrm{HM}}=S[\phi_{\max}]-S[\phi_{\min}]
= \frac{-24 \pi^2}{V_{\max}} + \frac{24 \pi^2}{V_{\min}}.
\end{equation} 
Action of an CDL instanton is the difference between the instanton solution and the homogeneous background (false vacuum),
 \begin{equation}\label{eq:CDLaction}
S_{\textrm{HM}}=S[\phi_{\textrm{CDL}}]-S[\phi_{\min}]
= S[\phi_{\textrm{CDL}}]+\frac{24 \pi^2}{V_{\min}}.
\end{equation}
We will describe calculation of $S[\phi_{\textrm{CDL}}]$ in detail in the following subsection \ref{sec:CDLinstantons}. 
However it is already clear that a CDL action is always smaller than the HM one because the instanton solution $\phi_{\textrm{CDL}}$ interpolates between the two vacua rather than constantly taking the highest value on top of the barrier as in HM solution. 

\begin{figure}[h]
\centering
\includegraphics[height=6.0cm]{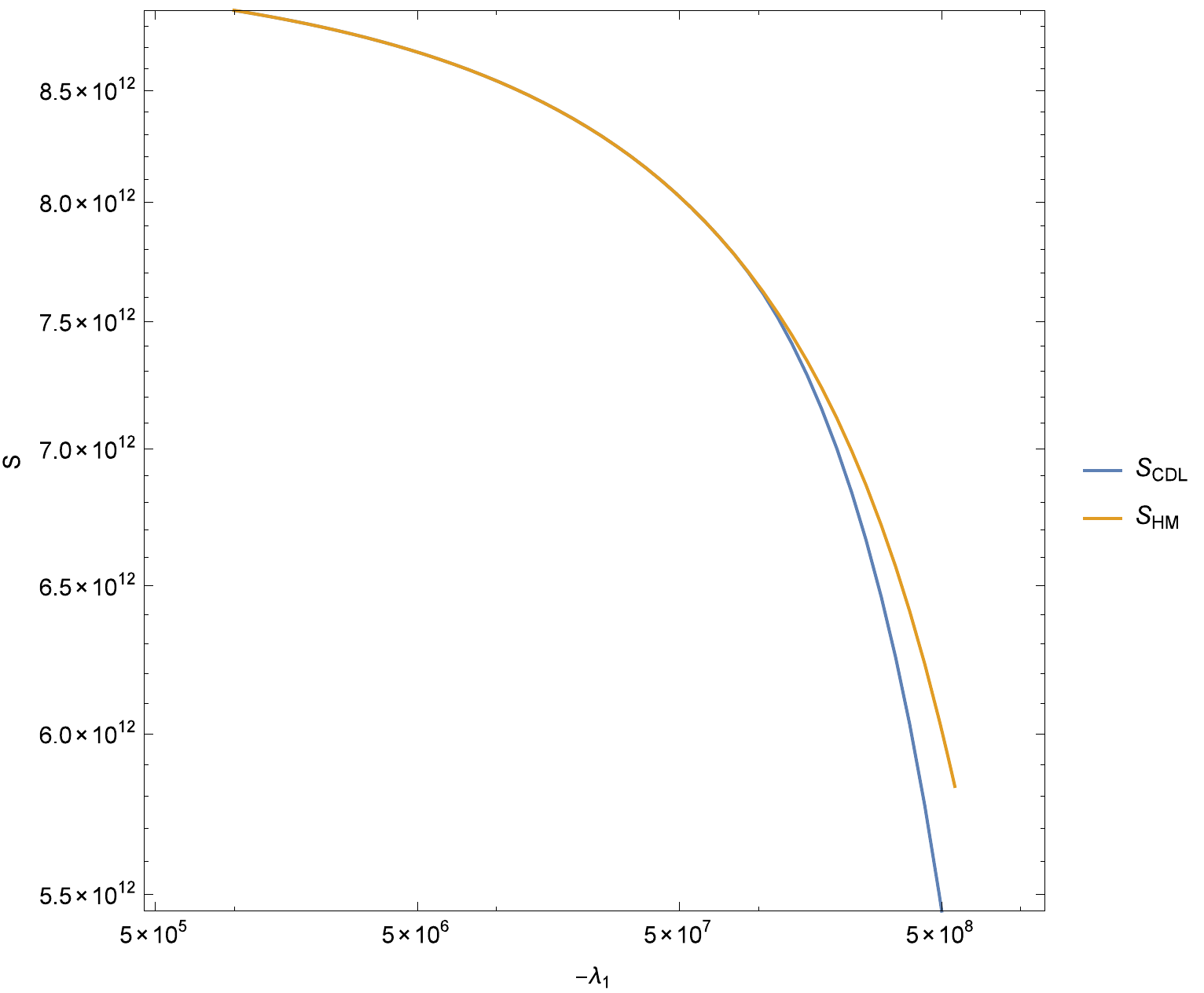}
\hspace{0.5cm}
\includegraphics[height=6.0cm]{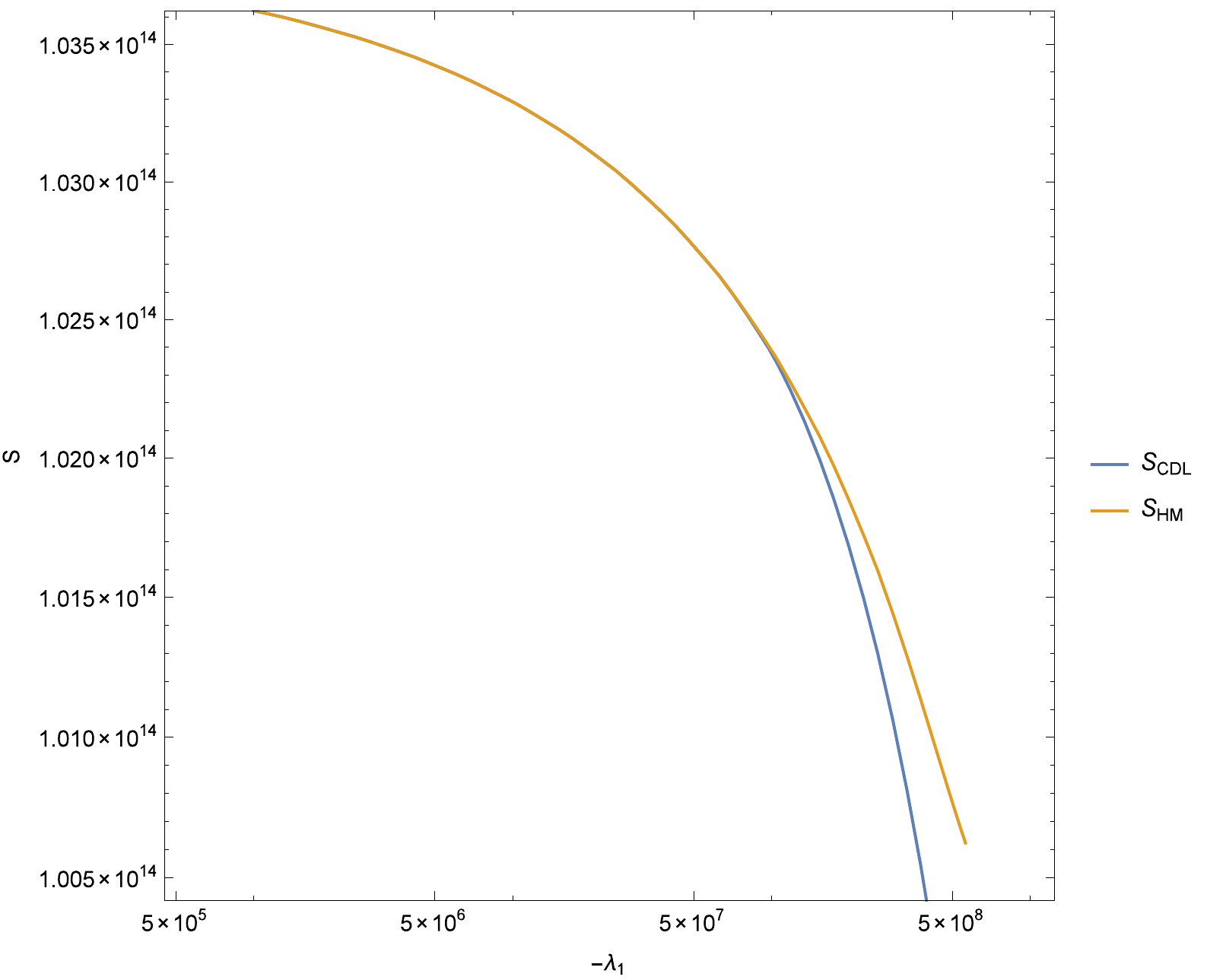}
\caption{\it Action of CDL and HM instantons with the false vacuum fixed at $V_{\min}=M^2/10$ (left panel) and $V_{\min}=M^2/100$ (right panel).} 
\label{fig:instantonaction}
\end{figure}

Figure~\ref{fig:instantonaction} shows actions of both solutions in our example potential. The resulting lifetime of the vacuum can be bounded from below by choosing the Planck scale as a dimensionfull quantity describing our potential,
\begin{equation}\label{eq:tau}
\tau =  \frac{3 H^3}{4 \pi M_p^4} e^{S}\approx 1.8 \times 10^{25} e^{S}s,
\end{equation} 
the current age of the universe is $T_U\approx 4.3 \times 10^{17} s$. Our result is huge compared to the current age of the universe despite Planck scale suppression, leaving absolutely no hope for a correct power spectrum after the tunneling finally occurs (as described in Section\ref{sec:evolofphi}). 

The only remaining loophole are oscillating bounces described in \cite{Hackworth:2004xb}. These bounces exist for values of $\beta$ between $1$ and $4$, and so they may dominate vacuum decay when CDL instantons do not exist. However when both CDL and HM solutions exist, actions of oscillating bounces are between the two. Thus looking at Figure~\ref{fig:instantonaction}, one immediatly sees from continuity of the solutions that the oscillating bounces are also completely irrelevant in our problem.

In summary we conclude that in the described potential the metastable vacuum is always significantly to long lived. And all solutions involving field being trapped in such a minimum due to classical evolution are excluded.   

\subsection{CDL instantons calculation}\label{sec:CDLinstantons}

We are interested in an $O(4)$ symmetric scalar field configuration $\phi=\phi(\tau)$, with the {Einstein frame} metric given by $ds^2=d\tau^2 + r(\tau)^2(d\Omega)^2$. Here $d\Omega$ is an infinitesimal element of the $3D$ sphere, and $r(\tau)$ is the {radius} of that sphere. The resulting metric tensor is of the form of the $FRW$ metric with the curvature parameter $k=+1$.
Euclidean action in Einstein frame takes the form
\begin{equation}\label{euclideanaction}
S_E=2\pi^2 \int d \tau r^3 \left( \frac{1}{2} \phi_{\tau}^2+V+\frac{1}{2} R \right)\, ,
\end{equation}
where $R=6\left( \frac{r_{\tau\tau}}{r}+\left( \frac{r_{\tau}}{r} \right)^2-\frac{1}{r^2} \right)$ and $\phi_{\tau} = \frac{d\phi}{d\tau}$. The equation of motion of the scalar field reads 
{\begin{equation}\label{EFEOM}
 \phi_{\tau\tau}+3\frac{r_{\tau}}{r} \phi_{\tau}=V_{\phi}\, ,
\end{equation}}
while the first Friedmann equation reads
\begin{equation}\label{EFFriedmanequation}
r_\tau=\sqrt{
1+ \frac{a^2}{3}\left(\frac{\beta}{4}\phi_{\tau}^2-V 
\right)}\, .
\end{equation}
One can show that scale factor $r$ crosses zero twice \cite{Guth:1982pn}. Without loss of generality we chose value of $\tau$ of the first one to be $\tau=0$, the other one we call $\tau_{\textrm{end}}$.
The appropriate boundary conditions then are
\begin{eqnarray}
\phi_{\tau}(0)=\phi_{\tau}(\tau_{\textrm{end}})=0 \\
r(0)=r(\tau_{\textrm{end}})=0 \nonumber
\end{eqnarray}
The final initial condition needed  for our equations is the initial field value $\phi_0$. For initial values $\phi_0$ to close the minimum of the potential the field will pass the other minimum, while for values too far from it it will start oscillating near maximum of the potential by the time when $r$ crosses its second zero. Between the two there is the correct value for which the field stops on the opposite slope of the barrier as $r$ crosses its second zero. We find this correct value corresponding to CDL  by a simple undershoot/overshoot method.

After finding the CDL solution for $\phi(\tau)$ and $r(\tau)$ we numerically perform the action integral in a form simplified using using equations of motion to finally get
\begin{equation}
S{[\phi_\textrm{CDL}]}=4\pi^2 \int d \tau \left( r^3 V(\phi(\tau)-3 r(\tau) \right).
\end{equation}
Which is the final result used in \eqref{eq:CDLaction}.

\section{Conclusions}\label{sec:concl}

In this paper we have analysed the evolution of the universe for the Starobinsky potential with higher order terms parametrised by two constants $\lambda_1$ and $\lambda_2$. In the Sec. \ref{sec:Infl} the general features of the Einstein frame potential have been discussed. Among them  the presence  of the plateau and of the steep slope, the existence of a minimum at $\phi_{\min}$ and of a local maximum at $\phi_{\max}$ (for $\lambda_1<-\lambda_s$) or a  saddle point for $\lambda_1 = -\lambda_s$. We have analysed the behaviour of the field near extreme points and established possible scenarios of the evolution of the system. In the special case of $\lambda_2 = 0$ the equivalent $f(R)$ theory which corresponds to the model considered in this paper has been found. It has been shown that higher order corrections to the potential weaken the problem of initial conditions of the Starobinsky inflation.
\\*

In the Sec. \ref{sec:infl2} the inflationary evolution of the system has been outlined together with the discussion of the power spectrum of initial inhomogeneities. The issue of eternal inflation has been addressed.  In fact, the existence of the steep slope has been proven to provide a solution to that problem for $\lambda_1 \gtrsim 10^{-3} $ or $\lambda_2\gtrsim 10^{-4}$. The attractor solution consistent with the slow-roll approximation has been determined. It is separated into two parts: one which leads to inflation on the Starobinsky plateau and one which leads to the minimum at $\phi_{\min}$. Those parts are separated by a local maximum $\phi_{\max}$, around which quantum fluctuations of $\phi$ dominate the evolution of the inflaton. This part of an attractor may be a source of topological inflation. We have found the maximal amount of e-folds which can be produced for general set of $\lambda$ parameters together with a peak of $N_{\max}$, which corresponds to the saddle point inflation.
\\*

In the Sec. \ref{sec:Temp} it has been proven that thermal corrections to the Einstein frame potential shall not modify significantly the evolution of the field. The inflaton trapped in $\phi_{\min}$ cannot leave the minimum via thermal correction because non-zero temperature can never (even for $T$ beyond the Planck scale) fill the minimum and push $\phi$ towards the GR vacuum. Temperatures larger than $M_{p}$ make the minimum even deeper than it is at  $T=0$.
\\*

In the Sec. \ref{sec:Tunnel} the issue of quantum tunnelling between the local minimum at $\phi_{\min}$ and the GR vacuum has been analysed. It has been proven that inflationary potentials which are able to generate $N_{\max}\geq60$ can generate tunnelling only via the Hawking-Moss effect, which gives a lifetime much bigger than the age of the Universe. For $\lambda_1\gg 10^6$, which is highly inconsistent with inflation, one obtains tunnelling via Coleman - De Luccia instantons and oscillating instantons, but the lifetime of the vacuum at $\phi_{\min}$ is again much bigger than the age of the universe. Thus the minimum at $\phi_{\min}$ is stable and $\phi$ can leave it only via the classical evolution.

\acknowledgments
This work was supported by the Foundation for
Polish Science International PhD Projects Programme co-financed by the EU
European Regional Development Fund.\\ 
This work was partially supported by National Science Centre, Poland 
under \\ research grants DEC-2012/04/A/ST2/00099 and DEC-2014/13/N/ST2/02712, and grant FUGA UMO-2014/12/S/ST2/00243.


\vspace*{0.3cm}

\begin{thebibliography}{99}

\bibitem{Starobinsky:1980te}
  A.~A.~Starobinsky,
  Phys.\ Lett.\ B {\bf 91} (1980) 99.
  
\bibitem{Ade:2013uln}
  P.~A.~R.~Ade {\it et al.}  [Planck Collaboration],
  Astron.\ Astrophys.\  {\bf 571} (2014) A22
  [arXiv:1303.5082 [astro-ph.CO]].


\bibitem{DeFelice:2010aj}
  A.~De Felice and S.~Tsujikawa,
  Living Rev.\ Rel.\  {\bf 13} (2010) 3
  [arXiv:1002.4928 [gr-qc]].
  
\bibitem{Codello:2014sua}
  A.~Codello, J.~Joergensen, F.~Sannino and O.~Svendsen,
  arXiv:1404.3558 [hep-ph].
  
\bibitem{Ben-Dayan:2014isa}
  I.~Ben-Dayan, S.~Jing, M.~Torabian, A.~Westphal and L.~Zarate,
  arXiv:1404.7349 [hep-th].

\bibitem{Artymowski:2014gea}
  M.~Artymowski and Z.~Lalak,
  JCAP09(2014)036
  [arXiv:1405.7818 [hep-th]].
  
  
\bibitem{Artymowski:2014nva}
  M.~Artymowski, Z.~Lalak and M.~Lewicki,
  arXiv:1412.8075 [hep-th].

  
  
\bibitem{Sebastiani:2013eqa}
  L.~Sebastiani, G.~Cognola, R.~Myrzakulov, S.~D.~Odintsov and S.~Zerbini,
  Phys.\ Rev.\ D {\bf 89} (2014) 023518
  [arXiv:1311.0744 [gr-qc]].
  
  
\bibitem{Motohashi:2014tra}
  H.~Motohashi,
  arXiv:1411.2972 [astro-ph.CO].

  
  
\bibitem{Kallosh:2013tua}
  R.~Kallosh, A.~Linde and D.~Roest,
  Phys.\ Rev.\ Lett.\  {\bf 112} (2014) 1,  011303
  [arXiv:1310.3950 [hep-th]].
  
  
\bibitem{Kallosh:2013yoa}
  R.~Kallosh, A.~Linde and D.~Roest,
  JHEP {\bf 1311} (2013) 198
  [arXiv:1311.0472 [hep-th]].
  
  
\bibitem{Kallosh:2014rga}
  R.~Kallosh, A.~Linde and D.~Roest,
  JHEP {\bf 1408} (2014) 052
  [arXiv:1405.3646 [hep-th]].
  
  
\bibitem{Kallosh:2014laa}
  R.~Kallosh, A.~Linde and D.~Roest,
  JHEP {\bf 1409} (2014) 062
  [arXiv:1407.4471 [hep-th]].
  
  
\bibitem{Galante:2014ifa}
  M.~Galante, R.~Kallosh, A.~Linde and D.~Roest,
  arXiv:1412.3797 [hep-th].
  
  
\bibitem{Broy:2014sia}
  B.~J.~Broy, D.~Roest and A.~Westphal,
  arXiv:1408.5904 [hep-th].
  
  
\bibitem{Kamada:2014gma}
  K.~Kamada and J.~Yokoyama,
  Phys.\ Rev.\ D {\bf 90} (2014) 10,  103520
  [arXiv:1405.6732 [hep-th]].
  
  
\bibitem{Guth:2007ng}
  A.~H.~Guth,
  J.\ Phys.\ A {\bf 40} (2007) 6811
  [hep-th/0702178 [HEP-TH]].
  
    
\bibitem{Ade:2013zuv}
  P.~A.~R.~Ade {\it et al.}  [Planck Collaboration],
  arXiv:1303.5076 [astro-ph.CO].
  
  
\bibitem{Ade:2014gua}
  P.~A.~RAde {\it et al.}  [BICEP2 Collaboration],
  arXiv:1403.4302 [astro-ph.CO].
  
  
\bibitem{Bojowald:2006da}
  M.~Bojowald,
  Living Rev.\ Rel.\  {\bf 8} (2005) 11
  [gr-qc/0601085].

  
\bibitem{Artymowski:2013qua}
  M.~Artymowski, Y.~Ma and X.~Zhang,
  Phys.\ Rev.\ D {\bf 88} (2013) 10,  104010
  [arXiv:1309.3045 [gr-qc]].
  
  
\bibitem{Zhang:2012em}
  X.~Zhang, Y.~Ma and M.~Artymowski,
  Phys.\ Rev.\ D {\bf 87} (2013) 8,  084024
  [arXiv:1211.4183 [gr-qc]].
  
\bibitem{Ijjas:2013vea}
  A.~Ijjas, P.~J.~Steinhardt and A.~Loeb,
  Phys.\ Lett.\ B {\bf 723} (2013) 261
  [arXiv:1304.2785 [astro-ph.CO]].
  
  
\bibitem{Kofman:2002cj}
  L.~Kofman, A.~D.~Linde and V.~F.~Mukhanov,
  JHEP {\bf 0210} (2002) 057
  [hep-th/0206088].
  

  
  
\bibitem{Bezrukov:2014ipa}
  F.~Bezrukov, J.~Rubio and M.~Shaposhnikov,
  arXiv:1412.3811 [hep-ph].
  
  
\bibitem{Linde:2005ht}
  A.~D.~Linde,
  Contemp.\ Concepts Phys.\  {\bf 5} (1990) 1
  [hep-th/0503203].
  
  
\bibitem{Kapusta:2006pm}
  J.~I.~Kapusta and C.~Gale,
  Cambridge, UK: Univ. Pr. (2006) 428 p

\bibitem{Hawking:1981fz}
  S.~W.~Hawking and I.~G.~Moss,
  Phys.\ Lett.\ B {\bf 110} (1982) 35.
  
\bibitem{Coleman:1980aw}
  S.~R.~Coleman and F.~De Luccia,
  Phys.\ Rev.\ D {\bf 21} (1980) 3305.
 
\bibitem{Hackworth:2004xb}
  J.~C.~Hackworth and E.~J.~Weinberg,
  Phys.\ Rev.\ D {\bf 71} (2005) 044014
  [hep-th/0410142].
  
\bibitem{Guth:1982pn}
  A.~H.~Guth and E.~J.~Weinberg,
  Nucl.\ Phys.\ B {\bf 212} (1983) 321.
  
\end{thebibliography}
\end{document}